\documentclass{article}
\usepackage{graphicx} % Required for inserting images
\usepackage{jheppub}
\usepackage{slashed}
\usepackage[T1]{fontenc}
\usepackage{braket}
\usepackage{multirow}
\usepackage[normalem]{ulem}
\usepackage{graphicx, graphics, hyperref, amsmath, amssymb, slashed, color, bbm,amsthm, array}

\subheader{\hfill \rm KEK-TH-2626}

\title{Energy Growth in $V_LV_L\to V_LV_L,\ V_LV_Lh$ Scattering to Probe Higgs Cubic and HEFT Interactions}

\author{Shameran Mahmud$^{1}$ and Kohsaku Tobioka$^{1,2}$}
\affiliation{\vspace{2mm} $^1$ Department of Physics, Florida State University, 77 Chieftan Way, Tallahassee, FL 32306, USA\\
$^2$Theory Center, High Energy Accelerator Research Organization (KEK), 1-1 Oho, Tsukuba, Ibaraki
305-0801, Japan}

\emailAdd{sm19bh@fsu.edu}
\emailAdd{ktobioka@fsu.edu}

\abstract{We compute the energy scales of perturbative unitarity violation in $V_L V_L \to V_L V_L h$ processes and compare them to $V_L V_L \to hhh$ process, where $V_L$ refers to a longitudinal mode of $Z$ or $W$ boson, and $h$ the Higgs boson. Using these energy scales, we determine which process is more sensitive to potential modifications in the Higgs sector at high-energy colliders.
Within the Higgs Effective Field Theory (HEFT), we consider the Higgs cubic coupling and other interactions with and without derivatives. Any HEFT interactions predict the perturbative unitarity violation at a finite scale, and in a generic case, the minimalistic process is $2\to 3$ scattering. 
Our analysis reveals that the energy scales for unitarity violation in $V_L V_L \to V_L V_L h$ and $V_L V_L \to hhh$ processes are similar across all scenarios considered. If the backgrounds are similar, $V_L V_L h$ final states are more feasible because $V_L V_L h$  has higher branching ratios in cleaner decay modes than $hhh$. 
We also investigate HEFT derivative interactions derived from various UV models. In these cases, both $V_L V_L \to V_L V_L$ and $V_L V_L \to hh$ processes exhibit unitarity violating behavior. We demonstrate that the energy scales for unitarity violation in $V_L V_L$ final states are comparable to or even lower than those in the $hh$ final state.
 }

\begin{document}
\maketitle
\flushbottom

\section{Introduction}
Since its discovery in 2012, the Higgs boson has been intensively studied and is considered a particularly intriguing and important particle due to its potential sensitivity to new physics beyond the Standard Model (BSM). A significant recent advancement in understanding the Higgs boson stems from Higgs coupling measurements, which involve assessing its couplings to fermions and gauge bosons through production and decay processes.
We observe that the SM is consistent with these measurements at an order of 10\% precision \cite{ATLAS:2022vkf, ATLAS:2024fkg, CMS:2022dwd, Mondal:2024ijr}, and the precision could be as low as a few percent at the High-Luminosity LHC (HL-LHC). This was possible due to efforts in the precision measurement of single Higgs production~\cite{Figy:2003nv, Spira:1995rr, Anastasiou:2016cez}.
On the other hand, these measurements do not probe the pure Higgs sector unless the Higgs boson has an unconventional kinetic term that modifies the Higgs couplings to fermions and gauge bosons. 

For the Higgs potential, the Higgs mass and the vacuum expectation value (VEV) were already well measured, but the other terms of the Higgs potential are poorly constrained. Therefore, probing the Higgs potential is essential. The current major effort along this line is to measure the Higgs cubic coupling, $ h^3$,  through the di-Higgs boson production at the LHC. 
The theory predictions for di-Higgs boson production include \cite{Baglio:2012np, Dolan:2012ac, Davies:2019dfy, Chen:2019lzz, Grazzini:2018bsd, Dreyer:2018qbw}. Since the cross section of this process is small, the current bound on the cubic coupling is at $\mathcal{O}(5)$ precision compared to the SM prediction~\cite{ATLAS:2021tyg, ATLAS:2023gzn, CMS:2022gjd, CMS:2022hgz}, and it will remain challenging at the HL-LHC, where the projected precision is $\mathcal{O}(50\%)$~\cite{Cepeda:2019klc}.

While measuring di-Higgs production is compelling, this observable alone is insufficient to pinpoint the underlying new physics. If a deviation is observed in the di-Higgs process, various modifications to the Higgs sector, beyond a simple shift in the $h^3$ coupling, could account for the data. Therefore, additional information from different processes is essential to narrow down the possibilities.

To conduct a general analysis, we employ the Higgs Effective Field Theory (HEFT) framework~\cite{Feruglio:1992wf, Bagger:1993zf, Koulovassilopoulos:1993pw}. 
The more common framework, the Standard Model Effective Field Theory ~(SMEFT)~\cite{Buchmuller:1985jz, Leung:1984ni, PhysRevLett.43.1566} with finite-dimensional truncation can be expressed by a finite expansion of the HEFT, but not vice versa. Another intriguing aspect of HEFTs which are not SMEFTs, is that the mass scale cannot be arbitrarily large, meaning that the high-energy colliders in the near future could probe the HEFT parameter space well. See Sec.~\ref{sec:HEFTintro} for more discussion. 

\begin{figure}[!tbp]
  \centering
  {\includegraphics[width=0.4\textwidth]{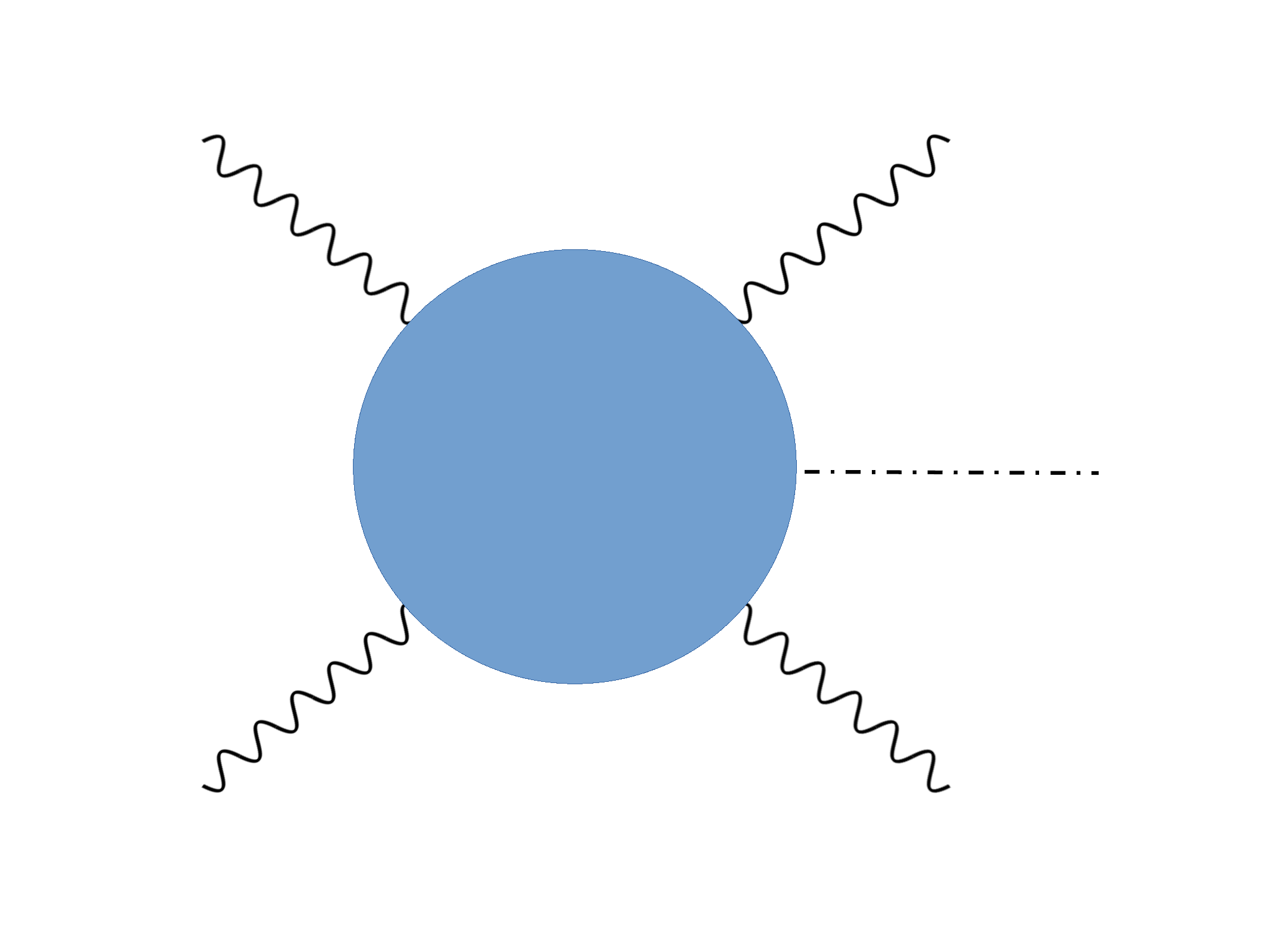}}
  ~{\includegraphics[width=0.4\textwidth]{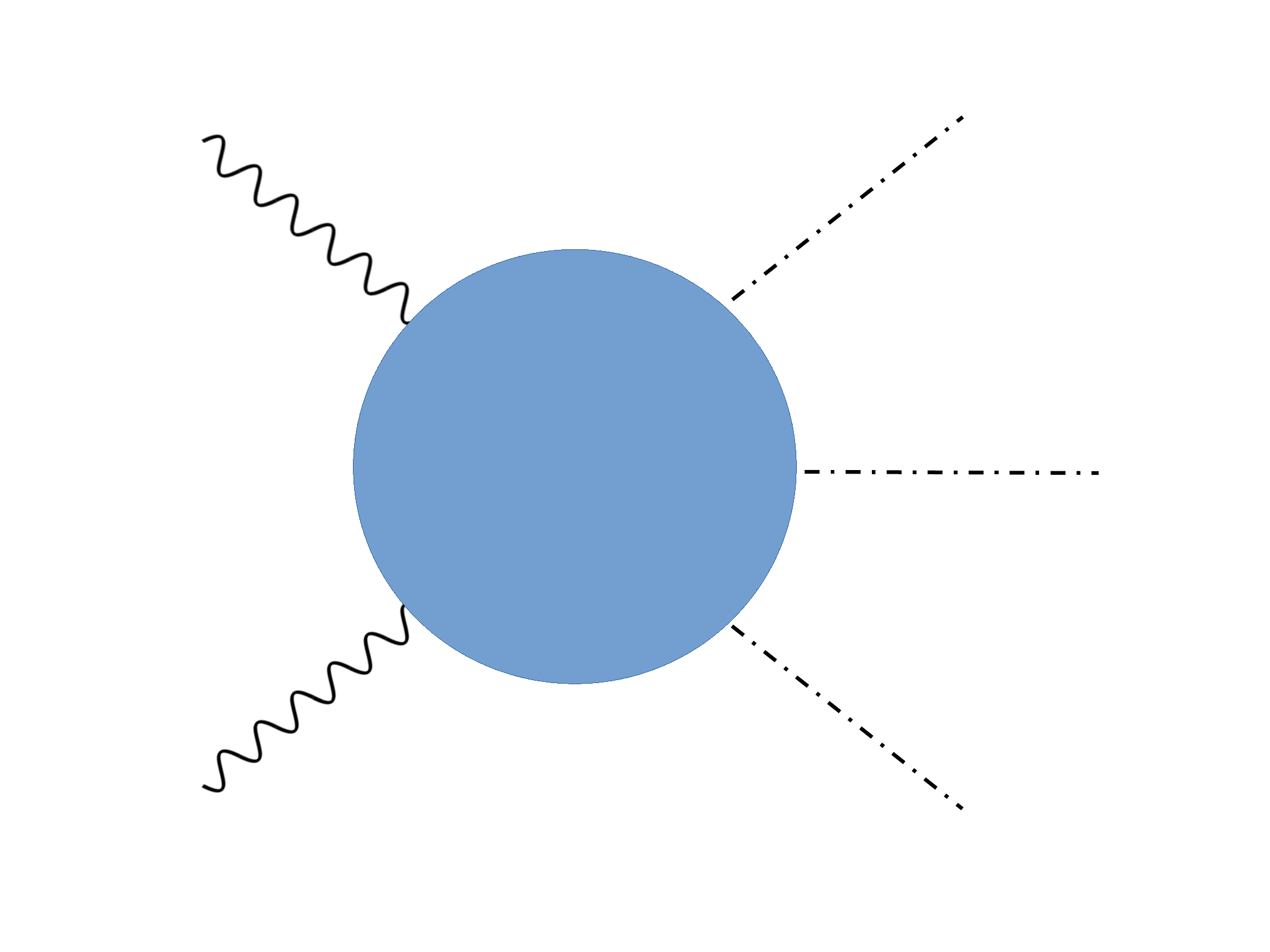}\label{fig:f12}}
  \caption{We study the $V_L V_L \to V_L V_L h$ process (left) and the $V_L V_L \to hhh$ process (right) within the HEFT, and show that the $V_L V_L \to V_L V_L h$ process has a similar order in the energy-growing behavior to the $V_L V_L \to hhh$ process.}
\end{figure}

Our main focus of this paper is to investigate the scattering processes of electroweak gauge bosons and Higgs boson based on the HEFT, in particular, the ones that have perturbative unitarity violation (from now on, called unitarity violation for brevity)  at high energy. 
It is well-known that the $2\to 2$ scattering of electroweak gauge bosons diverges at the high energy without the diagrams involving the Higgs boson because, in the SM, the scattering process in the electroweak sector has subtle cancellations involving the Higgs boson. 
 Similarly, almost any modification of the Higgs couplings can cause an incomplete cancellation, leading to unitarity violation at high energy. Therefore, the exploration of Higgs/gauge boson scattering processes is complementary to the di-Higgs measurement. 
Earlier works \cite{Falkowski:2019tft, Chang:2019vez} explicitly showed that the shift of the Higgs cubic coupling makes the $V_LV_L\to h^n$ scattering processes violate unitarity with $n\geq 3$,  where $V_L$ is a longitudinal mode of $W/Z$ boson, and $h$ is the Higgs boson. Another important feature is that higher collision energy will significantly enhance the cross section, which can be realized at the proposed FCC and muon colliders.

Regarding the unitary violating features of the HEFT, most of the previous works are presented using $V_LV_L\to~h^n$~\cite{Falkowski:2019tft, Chang:2019vez, Cohen:2021ucp, Gomez-Ambrosio:2022qsi, Delgado:2023ynh}. The equivalence theorem is applied in most of them, replacing the longitudinal gauge bosons ($V_L$) with the Nambu-Goldstone (NG) bosons ($G$) because the energy-growing behaviors are explicit. 
Exploiting the geometric features of HEFT, which are invariant under Higgs field redefinitions \cite{Alonso:2015fsp, Alonso:2016oah, Cohen:2020xca}, a calculation technique using sectional curvatures has also been established by Ref.~\cite{Cohen:2021ucp}. However, from the perspective of hadron collider searches, detecting multiple Higgs bosons is significantly more challenging than detecting multiple gauge bosons due to the large branching ratio of $h \to b\bar{b}$. This difficulty partly explains why di-Higgs boson searches have been arduous. Conversely, final states with multiple gauge bosons remain feasible. In fact, the ATLAS and CMS experiments have recently started observing triple gauge boson production \cite{ATLAS:2019dny, CMS:2020hjs}.

This paper argues that the final states with multiple gauge bosons often exhibit a similar strength of unitarity violation as in the fully Higgs boson final states. Roughly speaking, replacing Higgs boson pairs from the scattering process with longitudinal gauge boson pairs would maintain the unitarity violation, although there are exceptions for specific modifications, which are discussed in App.~\ref{app:replacing}.~\footnote{For example, if {\it only} the Higgs quartic coupling is shifted, the replacement does not work: $V_LV_L \to hhh$ has unitary violation while $V_LV_L \to V_LV_Lh$ does not.}  

For the demonstration, we mainly focus on the minimalistic processes, that is, the $2\to 3$ scattering as in Fig.~\ref{fig:f12}, and compare $V_LV_L \to V_LV_Lh$ to $V_LV_L \to hhh$ in the various HEFT scenarios including the shift of Higgs cubic coupling. The authors of \cite{Chang:2019vez} focused on the cubic coupling shift and showed that the unitarity violation appears in $Z_LZ_L \to Z_LZ_Lh$, $W_LZ_L \to W_LZ_Lh$,  and even in $V_LV_L \to V_LV_LV_LV_L$.
The importance of $2\to3(4)$ processes with multiple gauge bosons in the final state was also addressed in Refs.~\cite{Belyaev:2012bm, Abu-Ajamieh:2020yqi}. 
In our work, we consider various scenarios and comprehensively show which process will more effectively probe modifications of the Higgs potential than others. 
If the HEFT interactions involve derivatives, the $2\to 2$ scattering can violate unitarity, and we show that, considering several sets of HEFT interactions, $V_LV_L \to V_LV_L$ is at least comparable to  $V_LV_L \to hh$. The past works involving $2\to 2$ processes or derivative interactions in the HEFT include Refs.~\cite{Belyaev:2012bm, Abu-Ajamieh:2020yqi, Cohen:2021ucp, Kanemura:2021fvp, Garcia-Garcia:2019oig, Eboli:2023mny, Davila:2023fkk, Anisha:2024xxc}. The impact of HEFT operators through higher order calculations was studied, for example, in Ref.~\cite{Anisha:2024ljc}. The experimental search for the $V_L V_L$ final state has been performed, and it was observed at 2.3$\sigma$ at the CMS experiment~\cite{CMS:2020etf}. See also an ATLAS study for $q\bar q \to V_LV_L$~\cite{ATLAS:2023zrv}.

An intriguing feature of the HEFT is that the scale of unitarity violation is definite. In the previous works, the energy scale of unitarity violation, $E_*$, for the processes at very high multiplicities can be as low as (few) $\times 4 \pi v$  (i.e., more than 3-body final states). At low multiplicities, such as $2\to 3$ processes, the unitarity violation scales are still high, but our focus is not to demonstrate whether the $E_*$ values can be reached at the colliders. Instead, we use the $E_*$ values to rank the relevant processes.

Related works within the SMEFT \cite{Henning:2018kys, Chen:2021rid} also point out the unitary violation of $V_L V_L \to V_L V_L h/V_L V_L V_L$ processes. Note that the correspondence between the Higgs cubic coupling shift and the unitary violation rate would be slightly different from the HEFT case. For example, a SMEFT dimension-six operator, such as $(H^\dag H)^3$, will have different coefficients for the relevant contact terms compared to those from the HEFT operator shifting the Higgs cubic coupling, $(\sqrt{2 H^\dag H} - v)^3$. A relevant study is found in Ref.~\cite{Gomez-Ambrosio:2022why}. Also, the SMEFT operators generically do not predict unitary violation in arbitrarily high multiplicities, whereas the HEFT does.

The paper is organized in the following way. In Sec.~\ref{sec:HEFTintro},  we briefly review the difference between the HEFT and the SMEFT and highlight the relevant features of HEFT. Then, in Sec.~\ref{sec:PUV}, we discuss the perturbative unitarity violation of the gauge/Higgs boson scattering processes, in particular $V_LV_L \to V_LV_Lh$ and $V_LV_L \to hhh$, and briefly mention the experimental advantage of having multiple gauge bosons instead of Higgs bosons. 
 In Sec.~\ref{sec:Higgs-potential}, we present the results of the impact of HEFT operators, which only affect the Higgs potential, and in Sec.~\ref{sec:HEFTderivative}, we do the same for HEFT derivative operators motivated by several UV model EFTs.

\section{HEFT vs. SMEFT}\label{sec:HEFTintro}

In this paper, the main focus will be on interactions in the Higgs Effective Field Theory (HEFT) - in particular, the HEFT that \textit{do not admit} a SMEFT. The reason for this choice is explained in this section.

\subsection{Parametrizing the EFTs}
\begin{figure}[!tbp]
\centering
\includegraphics[width=350 pt]{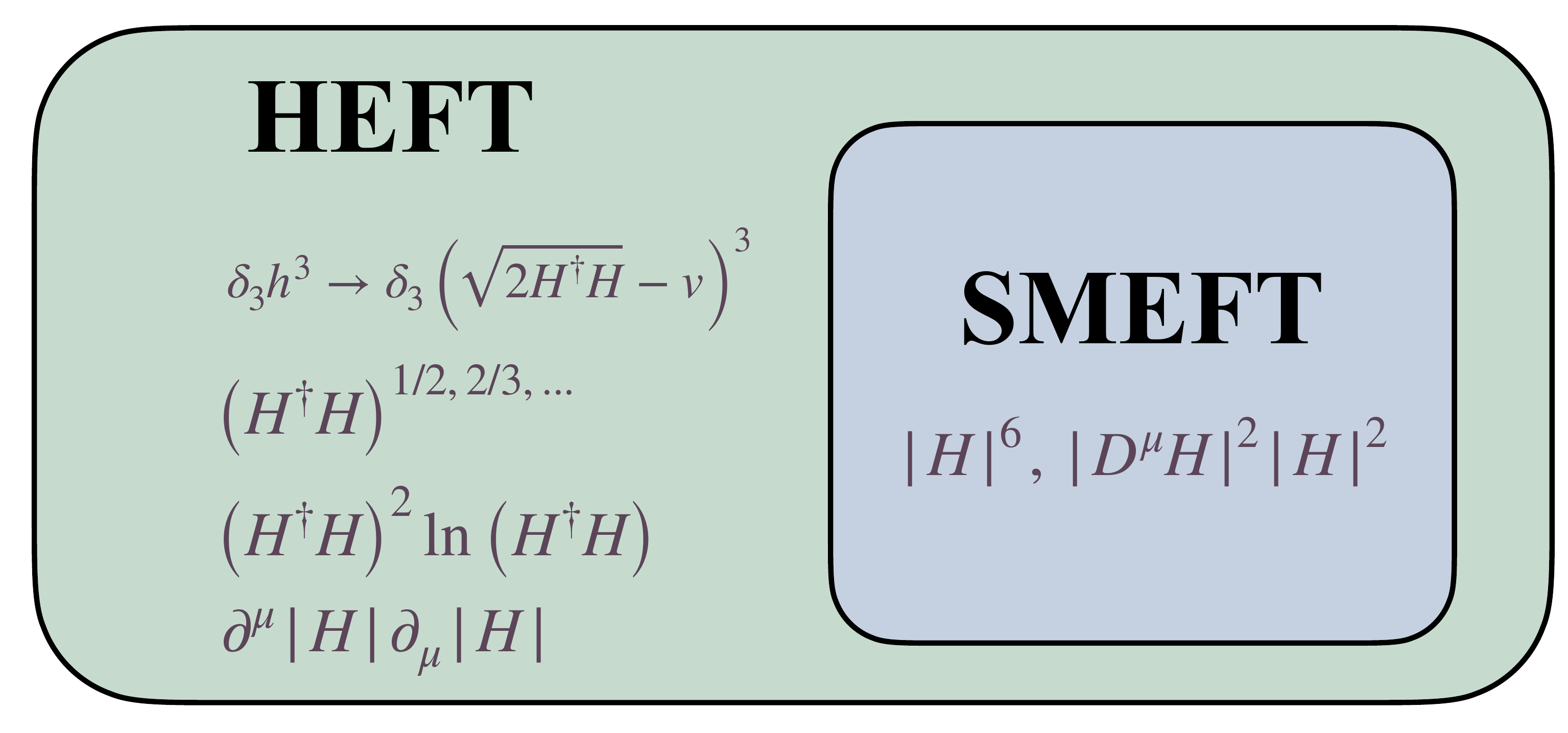}
\caption{Schematic diagram of all possible HEFTs. Every SMEFT, where the particles can be written as powers, is in the blue region; expanding $H^\dag H$ in terms of $h$ gives a HEFT parametrization. The green region (along with the boundary) contains theories that are purely HEFT. Here, writing $h$ in terms of $H^\dag H$ gives non-analytic terms in the Lagrangian. Those (here, referred to as ``HEFT'') will be the focus of this paper. }.
\label{fig:heftsmeft}
\end{figure}

In simplest terms, the difference between the two EFTs comes down to parametrization. In the Higgs sector, the SMEFT would be parametrized in terms of the $SU(2)$ Higgs doublet. One choice would be

\begin{equation}
\label{eq:su2}
    H = \frac{1}{\sqrt{2}}\left( \begin{array}c  \sqrt{2}G^+ \\ (v+h)+iG^0 \end{array}  \right),
\end{equation}
where $G^+, G^0$ are the NG bosons associated with spontaneous symmetry breaking, $v= 246$ GeV is the Higgs VEV, and $h$ is the singlet Higgs field. Using the Higgs doublet, a generic Higgs potential in the SMEFT parametrization is usually written in integer powers of $H^\dag H$,
\begin{equation}
\label{eq:smeft}
    V_{\rm SMEFT} = \sum_n \frac{c_n}{\Lambda^{2(n-2)}} \left(H^\dag H \right)^n.
\end{equation}
where the mass scale is $\Lambda$ and $c_n$ is a dimensionless coefficient. In the unitary gauge, the Higgs potential can also be given by the HEFT parametrization by using the singlet, $h$, in Eq.~\eqref{eq:su2},  
\begin{equation}
\label{eq:heft}
    V_{\rm HEFT} = \sum_\ell \tilde{c}_\ell h^{\ell},
\end{equation}
where $\tilde{c}_\ell$ is a dimensionful coefficient. The HEFT parametrization is more general, as different $\tilde{c}_\ell$'s are correlated to express the SMEFT potential of Eq.~\eqref{eq:smeft}  using $h$. 
Due to this, HEFT allows non-analytic potentials at $H=0$, which cannot be given by the SMEFT, for example, 
\begin{equation}
    V= g \left( H^\dag H \right)^\frac{2}{3} \supset  g v^\frac{4}{3} \sum_{n=3} \left( \begin{array}c \frac{4}{3} \\ n \end{array}  \right) \left( \frac{h}{v}\right)^n \ . 
\end{equation}
This illustrates the fact that the HEFT is a more general EFT than the SMEFT, as shown in Fig.~\ref{fig:heftsmeft}. It can encapsulate non-analyticities in the potential, which is described in more detail in Refs.~\cite{Alonso:2016oah, Falkowski:2019tft, Chang:2019vez, Cohen:2020xca}. 
Hereafter, when the HEFT is mentioned, it will be specifically about the space HEFT~\textbackslash SMEFT in Fig.~\ref{fig:heftsmeft}, i.e., the green region, while the SMEFT will refer to the blue region.

One caveat is that, with field redefinitions, some Lagrangian modifications may look non-analytic but are actually analytic. This is an issue that obfuscates whether or not some EFT is a SMEFT or a HEFT. This question has been discussed in depth in Refs.~\cite{Alonso:2015fsp, Alonso:2016oah, Cohen:2020xca, Gomez-Ambrosio:2022qsi}, and we recommend anyone with interest in this question to see the details in those references. 

\subsection{Why HEFT?}

Unitary violation exists in both SMEFT and HEFT. However, the HEFT has multiple reasons to be tested instead, from the unitary violation perspective. 

First of all, the SMEFT has an unknown mass scale, $\Lambda$, which can be as high as the Planck scale. The energy scale of unitarity violation, $E_*$,  is correlated to $\Lambda$, so there is no a priori reason that the unitarity violation can be seen even at future high-energy colliders. In the HEFT case, the energy scale of unitarity violation has to be definite, potentially as low as a few~TeV.~\footnote{As shown in Ref.~\cite{Cohen:2020xca}, the HEFT is induced by non-decoupling effects due to a new particle getting the majority of its mass from the Higgs mechanism. This leads to definite unitarity violation scales. Otherwise, the new particle can decouple by having a very high bare mass, leading to a SMEFT.} This means HEFT is a class of testable new physics. A well-known result shown in \cite{Falkowski:2019tft, Cohen:2021ucp} is that the $E_*$ value in the HEFT will be lower as the number of final states increases.

Another reason is that shifting a single Higgs potential term is covered by the HEFT rather than the SMEFT. The generic Higgs potential modification terms, as in Eq.~\eqref{eq:heft}, can be written in a gauge-invariant way as
\begin{align}
     h^\ell \to & \left( \sqrt{2 H^\dag H}- v\right)^\ell . 
\end{align}
A remark is that in di-Higgs measurements, the shift applied is only the cubic coupling, which would go under the category of HEFT. However, this difference is not important currently unless one includes NLO electroweak corrections of the di-Higgs production or looks at relevant higher multiplicity scattering processes which violate unitarity.
Given that a sizable coefficient for the cubic coupling is allowed by the current LHC constraints, let us show the expansion of the cubic coupling,
\begin{align}
\label{eq:h3}
 h^3 \to & \left( \sqrt{2 H^\dag H}- v\right)^3
  \supset h^3 + \frac{3 G^2}{2 v} h^2 \sum_{n=0}^\infty \left(- \frac{h}{v} \right)^n
  + \frac{3 G^4}{8 v^3} \sum_{n=0}^\infty (n+1) \left(- \frac{h}{v} \right)^n \left(2 v h- \frac{n+2}{2} h^2  \right)  \\\nonumber & +\frac{G^6}{16 v^5}  \sum_{n=0}^\infty \frac{(n+1)(n+2)}{2} \left(- \frac{h}{v} \right)^n \left\{ 2 v^2- (n+3)( 2vh - \frac{n+4}{4} h^2) \right\} + \cdots 
\end{align}
where $G^2=2 G^+ G^-+(G^0)^{2}$. This shift of cubic coupling leads to infinite NG boson-Higgs interactions. While the SMEFT also admits a cubic coupling shift, for example, by the dimension-six operator, 
\begin{equation}
    \left( H^\dag H \right)^3 = \frac{1}{8}((v+h)^2 +G^2)^3 \supset \frac{5}{2} v^3 h^3 ,  
\end{equation}
 there are only finite contact interactions between the NG bosons and Higgs.

Similarly, other HEFTs will also have infinite contact interactions due to expanding around $v$, meaning HEFTs, in general, have a lower $E_*$ value than SMEFTs. On top of this, any $V_L V_L \to h^n$ or $V_LV_L \to (V_L V_L)^n (h)$ process with a corresponding contact term will have unitarity violation in HEFT. Since collider searches prefer a lower multiplicity of final states, energy-growing processes that have the least amount of final states will be studied. In the next section, we show that in the cases with a potential modification, 3-body final states, particularly, the $V_L V_L \to hhh$ and the $V_L V_L \to V_L V_L h$ processes, will always have unitarity violation in HEFT. We also show that $2\to 2$ processes, that is, the $V_L V_L \to h h$ and the $V_L V_L \to V_L V_L$ processes, can also have energy growth based on specific HEFTs which admit derivative terms in the effective Lagrangian.

\section{Unitarity violation and $2\to 3$ scattering}\label{sec:PUV}
The modification of the Higgs sector leads to various consequences, one of which is the energy-growing behavior in scattering processes. The most well-known example is that the electroweak theory without the Higgs boson predicts the $V_LV_L \to V_LV_L $ scattering cross section grows with energy, and the (perturbative) unitarity seems to break down as the tree-level and higher-order contributions are comparable. 

We analyze tree level scattering processes of gauge/Higgs bosons as we look for perturbative unitarity violation. The equivalence theorem for NG bosons and gauge bosons is also used for simplicity, as gauge boson scattering would contain unitarity violation due to incomplete cancellations between multiple diagrams, while in the equivalence theorem picture, unitarity violation can occur due to a single contact term diagram.

\subsection{Averaged matrix element and $E_*$}
Here, we examine the impact of the HEFT interactions, including the Higgs cubic coupling, on the electroweak boson scattering processes. In order to quantify the unitarity violation, following Refs.~\cite{Falkowski:2019tft,Chang:2019vez, Cohen:2021ucp}, we extract the $s$-wave contribution by evaluating the phase-space averaged scattering matrix element, 
\begin{equation}
\label{eq:Mavg}
\hat{\mathcal{M}} = \left( \frac{1}{ \prod_i m_i! \prod_i n_i! \int \mathrm{dLIPS}_m \int \mathrm{dLIPS}_n}\right)^\frac{1}{2} \int \mathrm{dLIPS}_m \int \mathrm{dLIPS}_n \mathcal{M},
\end{equation}
 where $m$ is the number of ingoing particles, $n$ is the number of outgoing particles, and $m_i$ and $n_i$ are sets of indistinguishable particles in the initial and final state, respectively, so that $\sum_i m_i=m$ and $\sum_i n_i=n$, and $ \mathrm{dLIPS}_n = \prod_{i=1}^n \frac{\mathrm{d}^3p_i}{(2 \pi)^3 2 E_i}
 (2 \pi)^4 \delta^4\left(P-\sum_i p_i\right)$, where $P$ is the total momentum. 

To measure the strength of unitarity violation in each process, we find an energy scale, $E_*$,  at which the unitarity violates at the tree level, by checking the condition 
\begin{equation}
\label{eq:mhatcondition}
|\hat{\mathcal{M}}|^2 \simeq 1.
\end{equation}
The connection of unitarity violation with $\hat{\mathcal{M}}$ is discussed in App.~\ref{app:mhat}. The averaged matrix element is dimensionless, and for a constant $\mathcal{M}$, the corresponding cross section is obtained by roughly $|\hat{\mathcal{M}}|^2/E_{\rm cm}^2$ where $E_{\rm cm}$ is the center-of-mass energy, which is discussed in App.~\ref{app:x-sec}.  

Let us address two basic features of $E_*$: different scattering processes will have different $E_*$ values, and the process with lower $E_*$ is generally more sensitive to the underlying effects of new physics.  
The energy scale of unitarity violation indicates some new physics scale with a given modification of the Higgs sector, but new physics could cure the unitarity at a much lower scale, although we cannot a priori know this scale. This means $E_*$ is the highest possible scale for the new particle mass.

The main purpose of this paper is to sort out relevant processes in $2\to 3$ as well as $2\to 2$ scattering within various HEFT scenarios. Therefore, we use the $E_*$ values to rank some processes over others in terms of how sensitive they will be to new physics. This kind of information would be useful for experimentalists to set up programs for high energy gauge/Higgs boson scattering probes at the HL-LHC or other future colliders.   

\subsection{$V_L V_L \to hhh$ and $V_L V_L \to V_L V_L h$}
In this section, we elaborate on why vector-boson scattering to three bosons, in particular $V_LV_L h$ and $hhh$, will generally have energy-growing behavior. We also compare $V_L V_L h$ and $hhh$ final states from an experimental point of view. To be conservative, suppose only the Higgs potential is modified, namely, no new operators involving derivatives exist.   

Unitarity violation is easily understood using the equivalence theorem and the linear parametrization of the NG boson as in Eq.~\eqref{eq:su2}. Expanding the HEFT interaction by $h$ and $G$ gives almost arbitrary contact interactions (note that $G$ always appears as a pair). Each contact term leads to the constant contribution to the matrix element $\cal M $ of the $2\to n$ process, independent of $E_{\rm cm}=\sqrt{s}$. Non-contact contributions have negative powers of energy due to the propagators and, therefore, will not cancel the contact contribution. The averaged matrix element defined in Eq.~\eqref{eq:Mavg} has a phase space integral in the $n$-body final state, which gives energy dependence as  
\begin{equation}
    \int \mathrm{dLIPS}_n = \int \prod_{i=1}^n \frac{\mathrm{d}^3p_i}{(2 \pi)^3 2 E_i}
 (2 \pi)^4 \delta^4\left(\sum_i p_i\right) \simeq \frac{1}{8 \pi \left(n-1\right)! \left( n-2 \right)!} \left(\frac{E_{\rm cm}}{4 \pi} \right)^{2(n-2)}
\label{eq:LIPSn}
\end{equation}
where the last expression is in the massless limit. It is clear that for the $2\to 2$ processes, the constant matrix element does not lead to any energy-growing behavior in $\hat{\mathcal{M}}$; thus, no unitarity violation occurs. In contrast, $2\to 3$ or higher multiplicity processes predict unitarity violation as long as the constant matrix element exists. As expanding HEFT potential terms always gives contact interactions of high multiplicity, this behavior is guaranteed in HEFT potentials.

While the earlier works tend to focus on the $V_LV_L \to h^n$ processes with $n\geq 3$~\cite{Falkowski:2019tft, Chang:2019vez, Cohen:2021ucp, Gomez-Ambrosio:2022qsi, Delgado:2023ynh}, we would like to emphasize the $V_LV_L \to V_LV_L h$ process because the scale of the unitarity violation is often lower than $V_LV_L \to hhh$ process or at least comparable, as we will show in Secs.~\ref{sec:Higgs-potential} and \ref{sec:HEFTderivative}. 
Additionally, we point out that the experimental analysis would be easier than the multi-Higgs boson final states. 
In the multi-Higgs final states, extracting clean signatures would be more difficult than  $V_LV_L\to V_LV_L h$ because the Higgs boson decays dominantly to a bottom-quark pair, which suffers from the QCD background, and the clean decays of $h\to \gamma\gamma, \tau^+\tau^-, 4 \ell$ have small branching ratios. On the other hand, $V_LV_L h$ has more signal yield in the clean channels because the branching ratios of electroweak gauge bosons decaying to leptons is $\cal O$(10\%). 

\begin{table}
\centering
    \begin{tabular}{||c|c|c||}
    \hline
    final state $hAB$ & possible measured decays & branching ratio products \\ 
    \hline 
    \multirow{7}{0.3cm}{$hhh$} & ($b \bar b$)($b \bar b$)($\gamma \gamma$) & $ 2.08 \times 10^{-3}$ \\
    & ($b \bar b$)($b \bar b$)($\tau^+ \tau^-$) & $5.06 \times 10^{-2}$ \\ 
    & ($b \bar b$) ($\gamma \gamma$) ($\gamma \gamma$) & $8.27\times 10^{-6}$ \\  
    & ($b \bar b$)($\tau^+ \tau^-$)($\tau^+ \tau^-$) & $5.72 \times 10^{-3}$ \\  
    & ($b \bar b$)($\gamma \gamma$)($\tau^+ \tau^-$) & $4.77 \times 10^{-4}$ \\
    & ($\gamma \gamma $)($\tau^+ \tau^-$)($\tau^+ \tau^-$) & $2.70 \times 10^{-5}$ \\ 
    & ($\gamma \gamma $)($\gamma \gamma$)($\tau^+ \tau^-$) & $1.13 \times 10^{-6}$ \\ 
    \hline
    \multirow{2}{0.3cm}{$ZZ h$} & $ (b \bar{b}) (\ell^+ \ell^-) (\ell^+ \ell^-)$ & $5.41 \times 10^{-3}$ \\
     & $ (b \bar{b}) (q \bar{q}) (\ell^+ \ell^-)$ & $7.47 \times 10^{-2}$ \\
    \hline
    \multirow{2}{0.3cm}{$WW h$} & $ (b \bar{b}) (\ell \nu) (\ell \nu)$ & $5.62 \times 10^{-2}$   \\
     & $ (b \bar{b}) (q \bar{q}) (\ell \nu)$ & $ 0.233$   \\
    \hline
    \end{tabular}
    \caption{Examples of some branching ratios to compare a 3 Higgs boson final state to one with one Higgs boson and two gauge bosons. Higgs boson to $\gamma \gamma$ was chosen in this example as it is the cleanest channel available. Looking for multiple Higgs boson to $b \bar b$ decays is statistically difficult, as the $b$ pairs come in jets, which are much harder to analyze due to the difficulties of QCD. Also, processes involving $V_L$ also have a transverse background, which makes the signal gain more non-trivial. All the branching ratio values have been obtained from the PDG or the LHC Higgs working group. }
    \label{table:1}
\end{table}

\begin{figure}[!tbp]
  \centering
  {\includegraphics[width=0.4\textwidth]{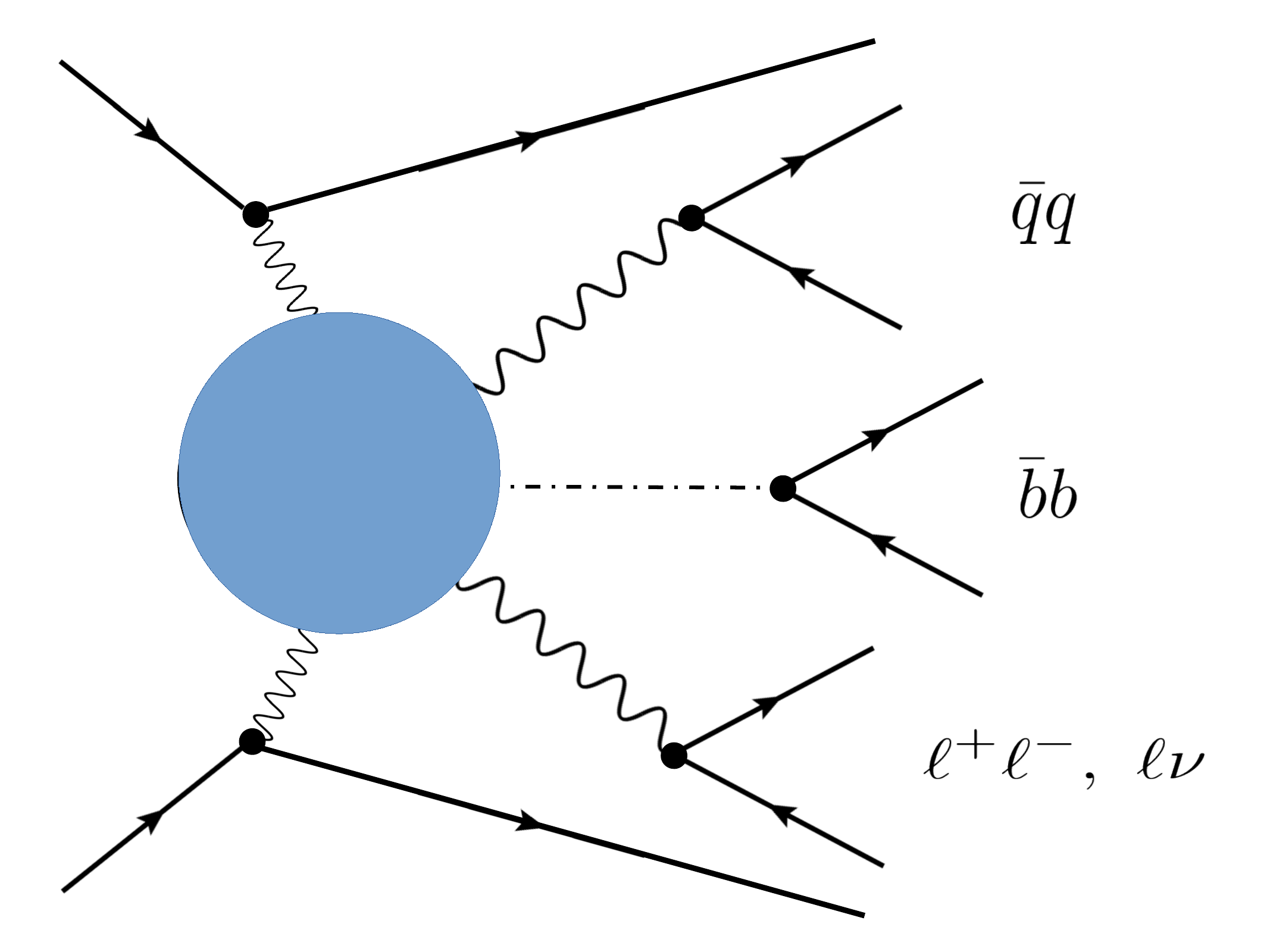}}~
 {\includegraphics[width=0.4\textwidth]{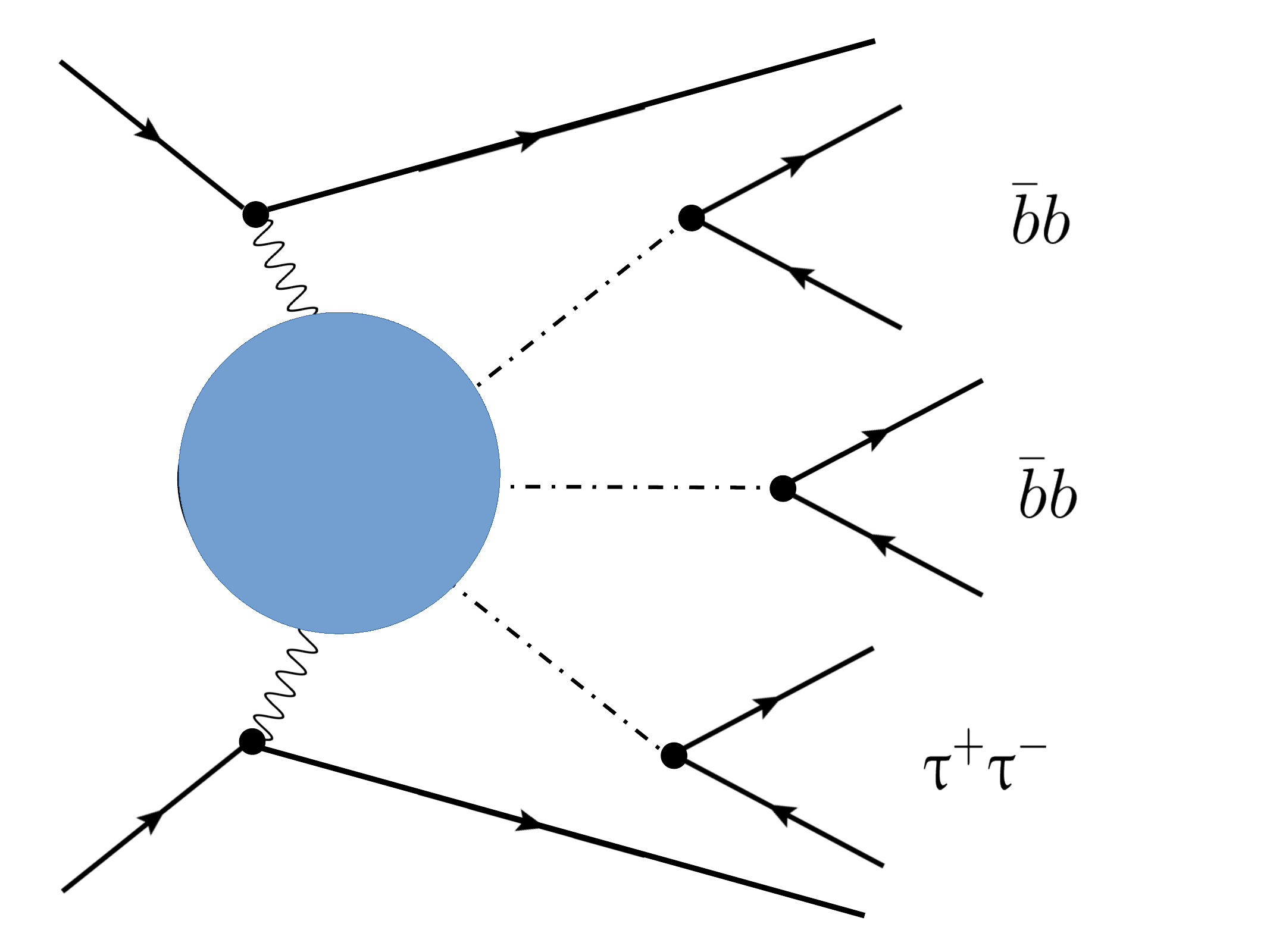}
  \caption{Most reasonable decay processes for both the $V_L V_L h$ and the $hhh$ final states at hadron colliders. The $V_L V_L h$ final state has the $V_L V_L$ decaying to $\bar{q} q $+$\ell \nu / \ell \ell$ and the $h$ decaying to a $\bar{b} b$ pair, compared to the $hhh$ final state where $hh$ decays to a $\bar{b} b\bar{b} b$ and the third $h$ decays to $\tau^+ \tau^-$. As shown in Table.~\ref{table:1}, the $V_L V_L h$ final state has a higher branching ratio than the $hhh$ final state with this choice of decay products.\label{fig:hhhVVh}}}
\end{figure}

In Table~\ref{table:1}, we show the products of branching ratios of Higgs and gauge bosons, assuming at least one Higgs boson decays to $b\bar b$; otherwise, a significant fraction of the signal would be lost. The representative processes are as in Fig.~\ref{fig:hhhVVh}. 
As we will show that the production cross sections would be similar,  it is more efficient to search for $W_LW_Lh$ as opposed to $hhh$ due to the branching ratios. For example, in decay modes with four jets, $WWh\to b\bar{b}q\bar{q}\ell\nu$ is larger than $hhh\to b\bar{b}b\bar{b}\tau^+\tau^- (b\bar{b}b\bar{b}\gamma\gamma)$  by a factor of 40 (100), and in modes with two jets, $WWh\to b\bar{b}\ell\nu\ell\nu$ is larger than $hhh\to b\bar{b}\tau^+\tau^-\tau^+\tau^- (b\bar{b}\gamma\gamma\tau^+\tau^-)$ by a factor of 10 (100). 

{Since the sensitivities depend on the background, we briefly discuss the possible backgrounds, although the detailed examination of the background is beyond the scope of this paper. 
For the di-Higgs boson search, the $hh\to b\bar{b}\gamma\gamma$ process is difficult at the LHC, because the continuum background ($\gamma\gamma$+jets) and single-Higgs background are  significant~\cite{ATLAS:2021ifb}, and the 
$hh\to b\bar{b}\tau^+\tau^-$ process is suffered from $t\bar{t}$, jets+$\tau_{\rm fake}$ and $Zb\bar{b}$ backgrounds~\cite{ATLAS:2022xzm}. If we naively extend this knowledge to the $pp\to hhhjj$ process, the continuum background would still be large for $hhh\to b\bar{b}b\bar{b}\gamma\gamma$. The di-Higgs boson processes could be a background for the triple Higgs boson signal. In the decay of $hhh\to b\bar{b}b\bar{b}\tau^+\tau^-$, we expect the background of $Z+b\bar{b}b\bar{b} / b\bar{b}jj$, $t\bar{t}b\bar{b}$, and jets+$\tau_{\rm fake}$. On the other hand, for the $pp\to V_LV_Lhjj$, the main background would be processes with two gauge bosons (without scattering each other) such as $hV_TV_Tjj$ and $V_TV_Tb\bar{b}jj$, which is inferred from the measurement of $W_LW_L\to W_LW_L$ at the CMS experiment~\cite{CMS:2020etf}. On top of this, Given that calculating the background processes at high multiplicities cannot be easily done, we focus on the signal yields in this paper.}

Let us comment on the other $2\to 3$ scattering processes, such as   $W_L W_L \to W_L Z_L W_L $ and $W_L W_L \to Z_L h h$. Their averaged-matrix element may grow with energy as well, despite having a propagator, which would bring $(E_{\rm cm})^{-2}$ dependence. However, their unitarity violations are not necessarily as strong as $V_LV_L h$ and $hhh$ since there are no diagrams from the contact term. Therefore, for simplicity and direct comparison to other references, the main focus in this paper will be $V_L V_L \to V_L V_L h$. Note, recently, the experimental search for triple gauge bosons has been performed~\cite {ATLAS:2019dny, CMS:2020hjs}. The unitarity violation in the $V_LV_LV_L$ final state is potentially interesting to constrain the Higgs couplings to the light quarks, top-quark, and electroweak gauge bosons~\cite{Abu-Ajamieh:2020yqi, Falkowski:2020znk}.

\begin{figure}[!tbp]
  \centering
  {\includegraphics[width=0.4\textwidth]{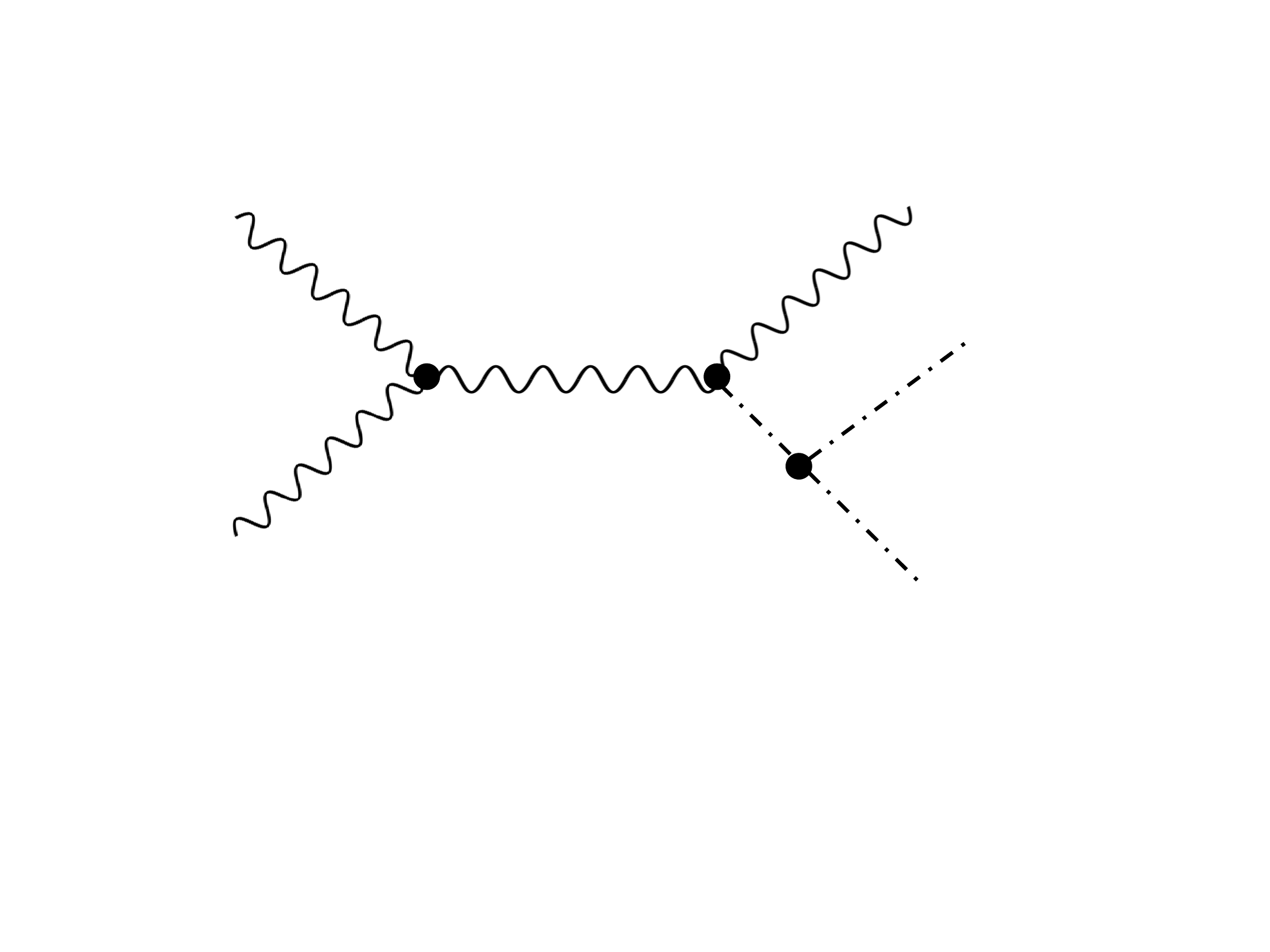}}
 ~{\includegraphics[width=0.4\textwidth]{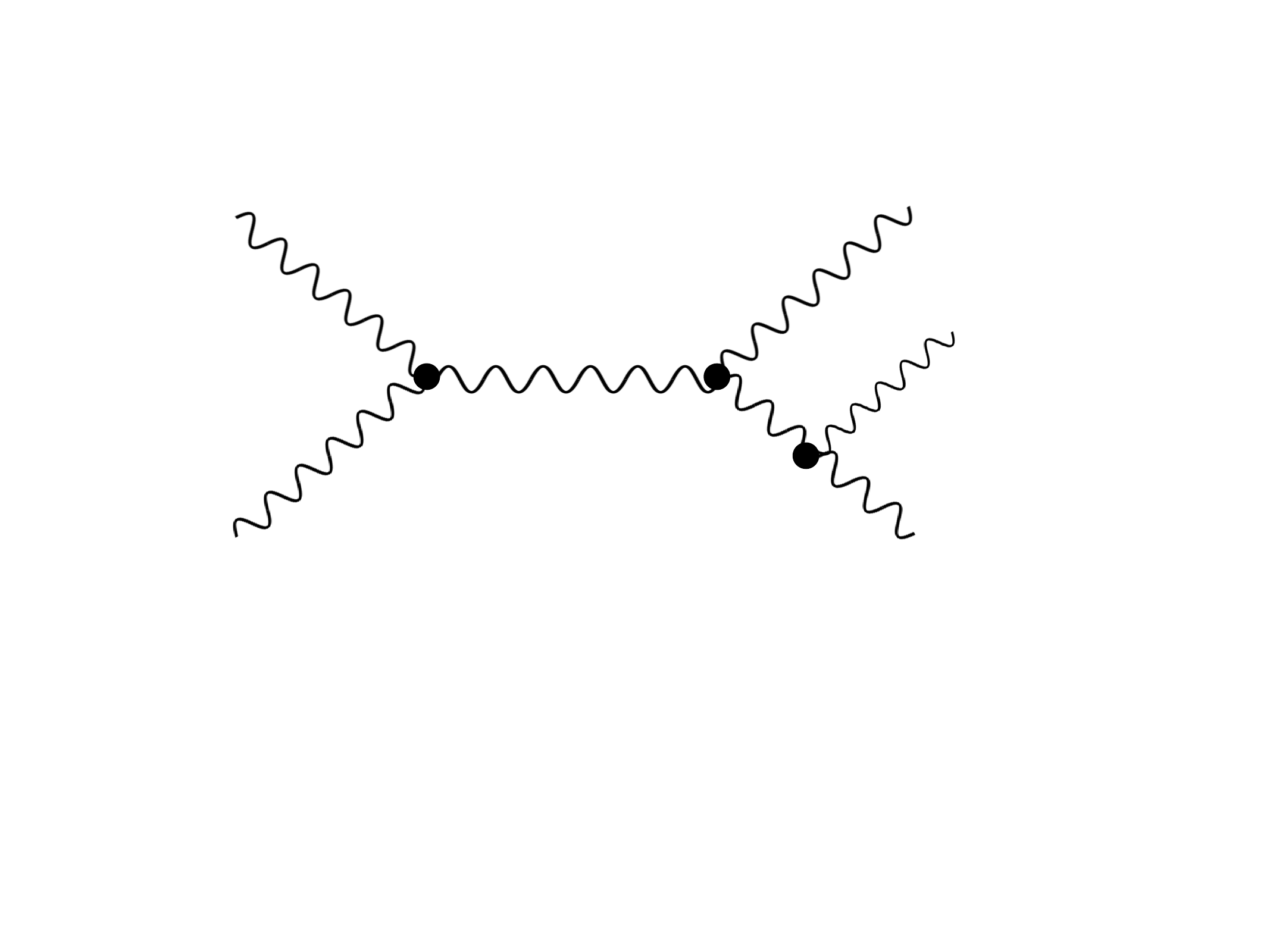}\label{fig:VVhVVV}}
 \vspace{-1.5cm} \caption{$V_L V_L \to V_L h h$ and $V_L V_L \to V_L V_L V_L$ processes may have energy-growing behavior due to new derivative interactions shifting the vertices. Because of the addition of propagators potentially reducing the energy growth, these are not studied in the paper.}
\end{figure}

\subsection{$V_L V_L \to hh$ and $V_L V_L \to V_L V_L$}

Until now, we have considered HEFT interactions without derivatives. If one considers UV models that can give a HEFT at low energy, the resulting set of HEFT interactions generally includes derivative terms that are as sizable as the potential terms. The derivative interactions introduce energy dependence into the contact interaction, adding energy-growing behavior to the matrix element $\cal M$. Therefore, the average matrix element of $2\to 2$ processes can violate the unitarity even if the phase space does not grow with energy. We investigate these processes in Sec.~\ref{sec:HEFTderivative} and see that the strength of energy growth in the $V_L V_L \to V_L V_L$ process is at least comparable to the one in the $V_L V_L \to hh$ process. Again, the experimental search for the two electroweak gauge bosons would be easier than the di-Higgs channel, as longitudinal gauge bosons have been produced at the LHC \cite{CMS:2020etf, ATLAS:2023zrv}. 
Such $2\to 2$ processes were also studied in Refs.~\cite{Cohen:2021ucp, Kanemura:2021fvp, Garcia-Garcia:2019oig, Eboli:2023mny}, which focuses on energy-growing behavior from derivative interactions in their Lagrangian.

\subsection{Connection to the SMEFT}
Note that when it comes to energy-growing behavior/unitary violation, the formalism in this paper can be applied to both SMEFT and HEFT. The HEFT was chosen for the reasons given in 
Sec.~\ref{sec:HEFTintro}. The two major differences are that in the case of SMEFT operators, the number of processes growing with energy is finite and that the $E_*$ values can be arbitrarily high because the new physics effect can decouple. Previous works on the SMEFT include  Refs.~\cite{Corbett:2014ora, Henning:2018kys, Chen:2021rid}.

\section{Unitarity violation for Higgs potential modifications}\label{sec:Higgs-potential}

We consider various interactions that affect only the Higgs potentials and investigate the relative importance of different scattering processes. One can view this as a more bottom-up approach because the EFTs from some UV models generically predict the HEFT interactions with derivatives, which is considered in Sec.~\ref{sec:HEFTderivative}. As discussed in Sec.~\ref{sec:Higgs-potential}, the HEFT interactions merely modifying the Higgs potential do not lead to the unitarity violation at the $2\to 2$ processes. Therefore, we focus on the $2\to3$ scattering processes.

Firstly, a generic Higgs cubic coupling shift is looked at, i.e., $V(h) \supset \frac{m_h^2}{2 v} \delta_3 h^3$. This was also discussed in Refs.~\cite{Falkowski:2019tft} and \cite{Chang:2019vez}, in the context of unitary violation in $V_L V_L \to h^n$ \cite{Falkowski:2019tft}, and more generally including a few NG boson processes, such as $Z_L Z_L \to Z_L Z_L h$ and $W_L Z_L \to W_L Z_L h$ \cite{Chang:2019vez}. In this paper, we also study other HEFT interactions as defined in Refs.~\cite{Cohen:2020xca, Cohen:2021ucp}. As mentioned in Sec.~\ref{sec:HEFTintro}, many HEFTs manifestly involve an infinite series of $\frac{h}{v}$ in all orders of the EFT expansion parameter. The Higgs cubic coupling shift, if it exists, would belong in the category of HEFT since writing $h^3$ in a gauge invariant manner gives 

\begin{equation}
h^3 = \left(\sqrt{2 H^\dag H}-v\right)^3\supset3 v^2 \sqrt{2 H^\dag H} + \left(2 H^\dag H\right)^\frac{3}{2}.
\end{equation}

Also, for all models, Custodial symmetry is assumed for simplicity; in general, HEFT can also violate custodial symmetry.

\subsection{Other Higgs potential modifications in HEFT}

Aside from a direct cubic coupling shift, we check a few other potential modifications in a framework of HEFT,

\begin{gather}
    V \supset \frac{1}{4 \pi^2} \left( H^\dag H\right)^2  \ln \left( H^\dag H \right), \\
   V \supset  \left( H^\dag H\right)^\frac{2}{3}, \\
    V \supset \sqrt{H^\dag H}.  
\end{gather}

The first potential commonly arises when integrating a heavy particle out at one loop, explaining the $ 4 \pi^2$ factor. The other two potentials are obtained at different parameter spaces after integrating out a second $SU(2)$ scalar doublet at the tree level. For example, $\sqrt{H^\dag H}$ is obtained in Ref.~\cite{Galloway:2013dma}. Usually, integrating out the new particle is also associated with non-analytic terms involving derivatives, which makes the theory grow faster. The $E_*$ values obtained from these terms will be checked in Sec.~\ref{sec:HEFTderivative}.

For the potential modifications, there are two ways to compute the $E_*$ value. Firstly, a linear parametrization is used, with the non-analytic potential expanded up to the needed number of NG bosons, for example,

\begin{align}
\label{eq:log}
\left( H^\dag H \right)^2 \ln\left( H^\dag H \right) \supset & \frac{v^4}{2} \ln\left( 1+ \frac{h}{v} \right) + \frac{v^4}{4}\ln\left(1+\frac{G^2}{\left(v+h \right)^2}\right) \\ \supset & v^2\frac{G^2}{4} \sum_{n=0}^\infty \left( n+1 \right) \left(-\frac{h}{v}\right)^n + 
\frac{G^4}{8} \sum_{n=0}^\infty \frac{\left( n+1 \right) \left( n+2 \right) \left( n+3 \right)}{6} \left(-\frac{h}{v}\right)^n + \cdots \nonumber
\end{align}
For this paper, we truncate the expansion at $\mathcal{O}(G^4h)$, as the $V_L V_L \to V_L V_L$ $(h)$ scattering is the main concern. If a process with  $V_L^6$ was wanted, the truncation should be done a term higher, and so on. 

Another approach, following \cite{Cohen:2021ucp}, involves a covariant and geometric approach, which uses the non-linear parametrization, so that $H^\dag H = \frac{(v+h)^2}{2}$. We check that both linear and non-linear parametrizations gave consistent results. One thing of note is that we exclusively used non-linear parametrization for the calculations in the Sec.~\ref{sec:HEFTderivative} following Ref.~\cite{Cohen:2021ucp}.

After obtaining a matrix element, $\mathcal{M}$, the momentum was averaged over using Eq.~\eqref{eq:Mavg}. The massless limit was taken for the potential case since such low multiplicities in the final states will have a much higher $E_*$ value than the mass of the $Z$, $W$, or $h$.

\begin{figure}[!tbp]
\centering
\includegraphics[width=450 pt]{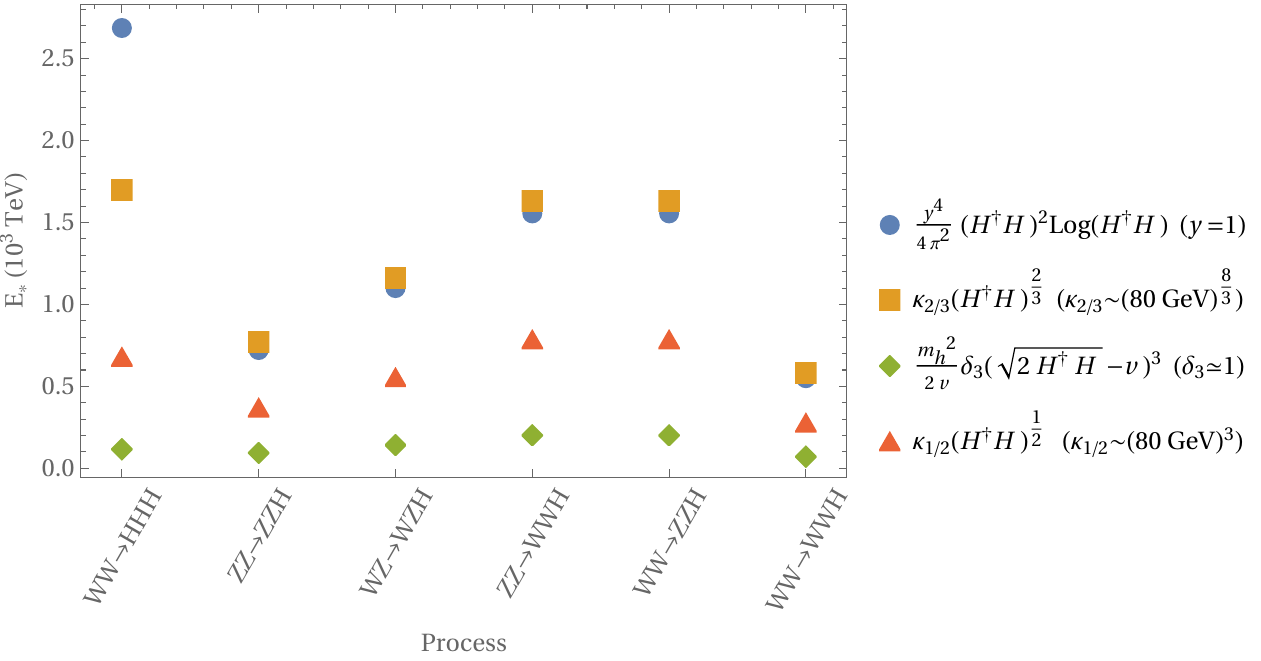}
\caption{Plot of $E_*$ values of multiple $2\to 3$ processes. As discussed in Sec.~\ref{sec:Higgs-potential}, four different interactions are considered. 
The first three processes for $(1+\delta_3)h^3$  were also studied in Ref.~\cite{Chang:2019vez}. The factor in front of the logarithm is due to the loop-level potential modification. The $E_*$ values in the $V_L V_L h$ final states are usually the same as, or better than, those in the $hhh$ final state, which does not depend on the choices of parameters. This is most apparent for the logarithmic term (blue circle).
\label{fig:res1}
}
\end{figure}

\subsection{Finding $|\hat{\mathcal{M}}|^2$ for the potential modifications}

For a potential modification only, the matrix element $\mathcal{M}$ can be found either by the standard QFT methods after expanding the potential or by taking the covariant derivative of the potential in the non-linear parametrization, as done in \cite{Cohen:2021ucp}.~\footnote{Note that here, the convention of \cite{Cohen:2021ucp} is followed. This gives an extra minus sign in $\mathcal{M}$.} Both methods will give the same result, but using the covariant derivative is more straightforward and can be generalized more easily for a given potential. For $V_L V_L  \to h^3$, 

\begin{equation}
\label{eq:vvpot}
    \mathcal{M}_{V_L V_L \to h^3}= \partial_h^3 \left(  f(h) \partial_h V(h)\right)|_{h=0},
\end{equation}
where $f(h) = \frac{1}{v+h}$. And for $V_L V_L \to V_L V_L h$, in the $(G^+, G^-, G^0)$ basis,
\begin{gather}
\label{eq:vvhpot}
    \mathcal{M}_{V_L V_L \to V_L V_L h} =
    \frac{1}{5} T_{i j k l} \partial_h \left\{3 f(h) \partial_h \left(f(h) \partial_h V(h) \right) - \partial_h^2 f(h) \partial_h V(h) - 2 \partial_h f(h) \partial_h^2 V(h) \right\}|_{h=0}.
\end{gather}
where $T_{i j k l} = \left( \sigma_{ij} \sigma_{kl} + \sigma_{ik} \sigma_{jl}  \sigma_{il} \sigma_{jk}\right)$, and

\begin{equation}
\label{eq:sigma}
    \sigma = \left(\begin{array}{c  c  c} 0 & 1 & 0 \\ 1 & 0 & 0 \\ 0 & 0 & 1 \end{array} \right).
\end{equation}
The main energy-growing contribution for both processes comes from the contact interaction terms in Eqs.~(\ref{eq:vvpot}, \ref{eq:vvhpot}), so for the purposes of finding the $E_*$ value, the contact terms are the only ones that need to be accounted for, as other terms just add corrections to it.

Given that the contact interactions are constant, the only non-trivial part is evaluating the phase space integral. Eq.~\eqref{eq:LIPSn} can be used as a good approximation of the integral, making obtaining $\hat{\mathcal{M}}$ straightforward. The details for the phase space integral are in App.~\ref{app:dlips}. 

\subsection{Results for the Higgs potential modifications} 

After doing the (tree level) analysis of the potential-only modifications, we find that the $V_L V_L \to V_LV_L h$ processes have similar $E_*$ values to the $V_L V_L \to hhh$ processes. The results are shown in Fig.~\ref{fig:res1}, and the tabulated values are given in Table~\ref{table:res1} of App.~\ref{app:values}. In the case of the logarithmic potential modification, all the $V_L V_L h$ final states have lower $E_*$ values. In other cases, some of the $V_L V_L h$ final states have lower $E_*$ values than the $hhh$ final state. For example, in the case of the fractional potential modifications, $(H^\dag H)^\frac{1}{2}$ and $(H^\dag H)^\frac{2}{3}$, the $W_L W_L \to W_L W_L h$ and the $Z_L Z_L \to Z_L Z_L h$ processes have lower $E_*$ values than the $W_L W_L \to hhh$ process. 

While the $E_*$ values are too high to be accessed at the colliders, we focus on the minimalistic processes and demonstrate that the $V_LV_L\to V_LV_Lh$ processes and the $V_LV_L\to hhh$ processes have similar energy scales of unitarity violation in the various HEFT scenarios. The results for the Higgs cubic modification match what was shown in Ref.~\cite{Chang:2019vez}. 
In order to reach even lower $E_*$ values, we have to explore higher multiplicity processes beyond 3-body final states. One intriguing process is $V_L V_L \to V_L V_L V_L V_L$, and it is shown in Ref.~\cite{Chang:2019vez} that this process has significantly lower $E_*$ value and suitable to constraint the Higgs cubic coupling.

\section{Unitarity violation for the HEFT with derivatives}\label{sec:HEFTderivative}

Another case to evaluate the importance of multi-gauge boson final states is HEFT interactions involving derivatives. Here, instead of picking arbitrary operators, we investigate sets of HEFT operators motivated by viable UV models, which were studied in Ref.~\cite{Banta:2021dek}.
In this approach, to obtain derivative terms, three UV models were chosen. A heavy singlet scalar and a heavy vector-like fermion are chosen based on the viable loryon searches using Ref.~\cite[Fig.~8, Fig.~10]{Banta:2021dek}. These models admit a potential term similar to $\frac{1}{4 \pi^2} \left( H^\dag H\right)^2  \ln\left( H^\dag H \right)$.  On top of this, a model based on the two-Higgs doublet model, briefly mentioned in Ref.~\cite{Falkowski:2019tft}, was chosen, with parameters that will admit a $(H^\dag H)^\frac{2}{3}$ term in the potential. For this analysis, because the $E_*$ values obtained from the derivative terms are much lower than those from the potential terms due to the very strong energy-growing behavior, the main focus is on the derivative terms. These terms have energy dependence as well, even at the $2 \to 2$ level, so both the $2 \to 2$ and $2 \to 3$ processes are looked at.

The two models chosen from Ref.~\cite{Banta:2021dek} induce HEFTs via a loop. The first one is

\begin{equation}
    \mathcal{L}_{\mathrm{UV}, S}= \mathcal{L}_{\mathrm{Higgs}}-\frac{1}{2}S \left( \partial^2+m^2+\lambda H^\dag H \right) S, 
\end{equation}
where $S$ is a singlet under $SU(2)$, and $\mathcal{L}_\mathrm{Higgs}$ is the standard model Higgs sector Lagrangian. The other EFT is a set of vector-like fermions where a non-zero bare mass is allowed,
\begin{equation}
 \mathcal{L}_{\mathrm{UV}, \Psi}= \mathcal{L}_{\mathrm{Higgs}}+ \bar \Psi \left( i \slashed{D} - m - \Xi \right) \Psi,
\end{equation}
 where 
 \begin{equation}
     \Psi = \left( \begin{array}{c} \psi \\ \chi   \end{array} \right),
 \end{equation}
and $\psi$ and $\chi$ are Dirac fermion doublets for which the handed components transform under the SM $SU(2)_L$ and $SU(2)_R$ respectively, and 

\begin{equation}
    \Xi = \left( \begin{array}{c c} 0 & y \xi \\ y^* \xi^\dag & 0 \end{array}\right),
\end{equation}
where $\xi = \left( \begin{array}{c c} i \sigma_2 H^* & H \end{array}\right)$. These choices were made to enforce Custodial symmetry in the product $y \bar \psi \xi \chi + (\rm h.c.)$ and to have a non-zero gauge invariant bare mass. In this paper, the bare mass is turned off. One thing to note is that the fermion model, by itself, actually breaks $\kappa_\gamma$ constraints. However, this could be fixed by adding other BSM particles, which may bring $\kappa_\gamma$ to the allowed range. 

Another (more constrained) model, briefly discussed in \cite{Falkowski:2019tft}, is the addition of a scalar $SU(2)$ doublet, $\Sigma$, with no bare mass,

\begin{equation}
    \mathcal{L}_{\mathrm{UV}, \Sigma}= \mathcal{L}_{\mathrm{Higgs}}+|D \Sigma|^2+\kappa^2 (\Sigma^\dag H + \mbox{h.c.})-\lambda_\Sigma (\Sigma^\dag \Sigma)^2.
\end{equation}
We choose one viable benchmark point per model. Additionally, we include one more viable benchmark point for the fermion to demonstrate that the $hhh$ final state has different coupling scaling on $E_*$ compared to the $V_L V_L h$ final state. For the SU(2) doublet addition, one benchmark point based on the allowed ranges of the Higgs cubic coupling shift was chosen. From these benchmark points, we obtain sets of HEFT interactions and calculate the $E_*$ values for each one. 

To make sure to be in the parameter space where the HEFT-like behaviors are the strongest, the bare mass of all of the loop theories are set to $0$ as well. These theories still receive a mass from the Higgs. 

In this case, since the unitarity violation scales for the $V_L V_L h$ final states had momentum and angular dependences, the full phase space integral was done without the approximation in Eq.~\eqref{eq:dlips}. The integral was not simply $\int \mathrm{dLIPS}_3$, but $\int \mathrm{dLIPS}_3 \mathcal{M}$ . Details are given in Sec.~\ref{subsec:mder} and App.~\ref{app:dlips}.

\subsection{Obtaining the EFT}

As stated in \cite{Cohen:2020xca}, for HEFT, the EFT must be all orders in the $h$ and two orders in the derivative.
Suppose there exists a heavy new particle, $\phi$. The simplest way to find the lowest order EFT for $\phi$ is by using tree-level matching. This is done by minimizing the potential for $\phi$, and hence obtaining $\phi[H]$, where $H$ is the Higgs field. Due to the derivative terms for $\phi$, the effective Lagrangian will naturally contain the 2nd order derivative terms. For example, in the case of the scalar doublet addition,

\begin{equation}
    \frac{\partial V_\Sigma}{\partial \Sigma^\dag} = \kappa^2 H - \lambda_\Sigma (\Sigma^\dag \Sigma) \Sigma = 0 \to \Sigma[H]= \left( \frac{\kappa^2}{2 \lambda_\Sigma} \right)^{\frac{1}{3}} \frac{H}{\left(H^\dag H \right)^\frac{1}{3}}.
\end{equation}
Then, the effective Lagrangian reads

\begin{equation}
    \mathcal{L}_{\mathrm{EFT}, \Sigma} = \mathcal{L}_{\mathrm{Higgs}}+\left( \frac{\kappa^2}{2 \lambda_\Sigma} \right)^\frac{2}{3} \frac{ \left|D H\right|^2 - \frac{8}{9}  \left(\partial \sqrt{H^\dag H}\right)^2}{\left(H^\dag H \right)^\frac{2}{3}}+\frac{3}{4} \left( \frac{\kappa^8}{\lambda_\Sigma} \right)^\frac{1}{3} \left( 2 H^\dag H \right)^\frac{2}{3}.
\end{equation}
Using $H= \frac{v+h}{\sqrt{2}} U$ gives the HEFT interactions.

Using the effective potential formalism expanded to 2 derivative order gives us the loop EFT, in the same way as App. D in \cite{Cohen:2020xca} or App. A in \cite{Banta:2021dek}. The results are restated here in the $m=0$ limit for convenience:

\begin{equation}
    \label{eq:scalar}
    \mathcal{L}_{\mathrm{EFT}, S} = \mathcal{L}_{\mathrm{SM}}+\frac{\lambda}{96 \pi^2} \left(\partial \sqrt{H^\dag H}\right)^2+\frac{\lambda^2}{64 \pi^2} \left(H^\dag H \right)^2 \left(\ln \frac{\mu^2}{\lambda H^\dag H}+\frac{3}{2} \right),
\end{equation}
for the scalar and

\begin{equation}
    \label{eq:fermion}
     \mathcal{L}_{\mathrm{EFT}, \Psi} = \mathcal{L}_{\mathrm{SM}}+\frac{y^2}{4 \pi^2} \left(
     |D H|^2\ln \frac{\mu^2}{y^2 H^\dag H} -\frac{2}{3}\left(\partial \sqrt{H^\dag H}\right)^2\right)- \frac{y^4}{4 \pi^2}(H^\dag H)^2 \left(\ln \frac{\mu^2}{y^2 H^\dag H}+\frac{3}{2} \right),
\end{equation}
 for the fermion, where $y$ was chosen to be real for simplicity, and we choose $\mu=m_f = \frac{y v}{\sqrt{2}}$ to minimize the logarithm. Note that the scalar case in Eq.~\eqref{eq:scalar} does not have any $\mu$ dependence in the derivative terms.~\footnote{In fact, even with a logarithmic potential modification in the contact terms, $\mu$ is only important in the $2\to 2$ case.} As stated in the beginning of this section, the potential terms will be ignored for this analysis.

While \cite{Cohen:2021ucp} covered the scalar singlet model as well, different benchmarks were chosen, and it was in the context of multiple Higgs final states. Here, the main addition will be multi $(V_L V_L)^n$ $(h)$ final states (for now, $V_L V_L$).

\begin{figure}[!tbp]
\centering
\hspace*{-1.5cm}
\includegraphics[width=500 pt]{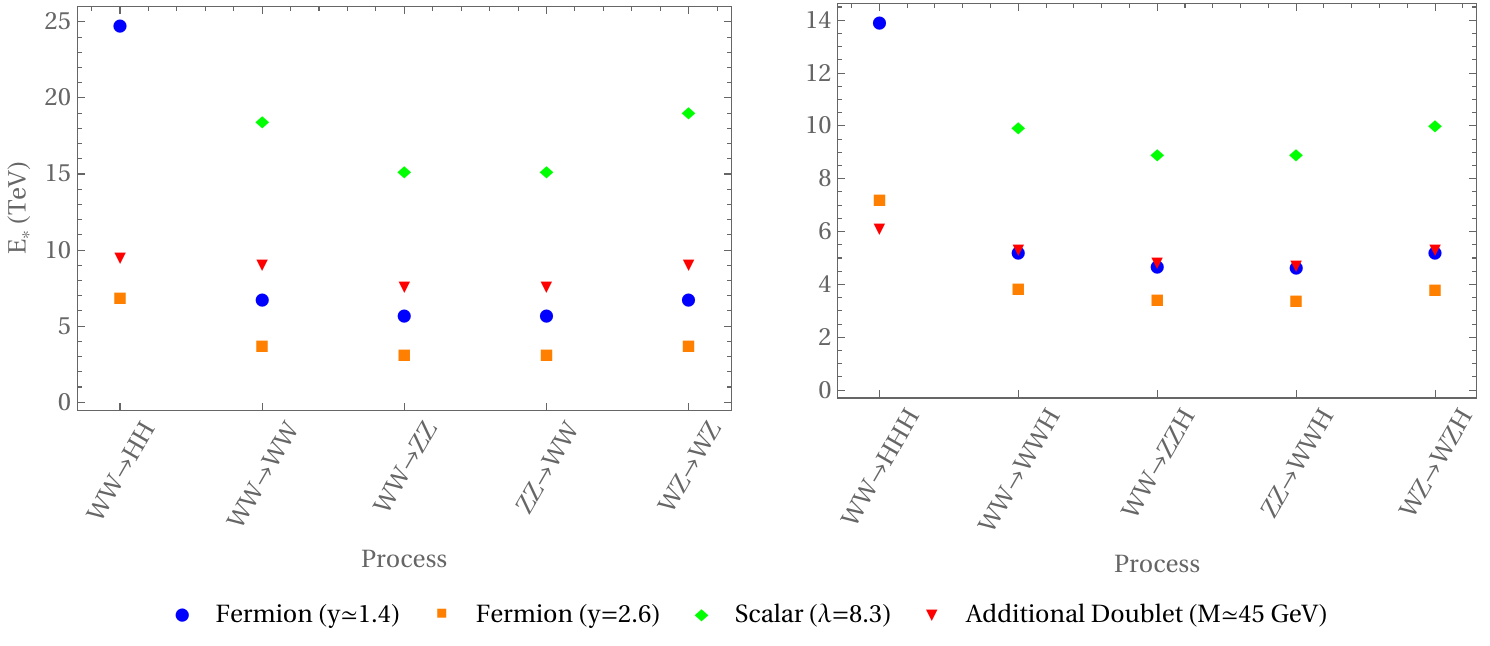}
\caption{Plots of $E_*$ values of multiple $2\to 2$ (left) and $2\to 3$ (right) processes. The HEFT interections involve derivatives, as discussed in Sec.~\ref{sec:HEFTderivative}. The  $V_L V_L \to h^n$ processes in the scalar case are studied in \cite{Cohen:2021ucp}, but we choose different parameters. Here, $M \equiv \left( \frac{\kappa^2}{\lambda_\Sigma} \right)^\frac{1}{2}$, and $\mu=m_f$, where $m_f$ is the mass of the fermion being integrated out. The $W_L W_L \to h^n$ processes had no $E_*$ values of $\mathcal{O}$(TeV) in the scalar case, as a bare mass is required for $\mathcal{K}_h|_{h=0}$ to be non-zero. Also, we adopt two benchmark points of the fermion case because the scaling in the couplings is different between the $hh(h)$ final states and the $V_LV_L(h)$ final states due to the differences in $\mathcal{K}_{h,\Psi} \propto y^4$ and $\mathcal{K}_{\pi,\Psi}\propto y^2$.} 
\label{fig:res2}
\end{figure}

\subsection{Finding $|\hat{\mathcal{M}}|^2$ for the derivative terms}\label{subsec:mder}

In this case, due to the derivative interactions, the derivative terms of the Higgs and NG bosons will no longer be canonically normalized and, in the non-linear parametrization, will look like

\begin{equation}
   \mathcal{L}_{\mathrm{kin}} = \frac{1}{2}K^2(h) \left(\partial h \right)^2+\frac{1}{2} [v F(h)]^2 \left| \partial U \right|^2.
\end{equation}
Here, the Higgs doublet, $H$, is written as $ \frac{v+h}{\sqrt{2}} U $, where $U= \exp(i \frac{T^a G^a}{v})  \left(\begin{array}{c} 0 \\ 1 \end{array} \right)$, with $T^a$ being the $SU(2)$ generators. $U$ encodes all the information on the NG bosons, and $h$ is the Higgs field. For example, in the case of the additional $SU(2)$ doublet,

\begin{equation}
\label{eq:doublet}
    \mathcal{L}_{\mathrm{kin},\Sigma} = \frac{1}{2}\left[1 + \frac{1}{9} \left( \frac{M}{v+h} \right)^\frac{4}{3} \right] \left( \partial h \right)^2 + \frac{1}{2}\left[1 + \left( \frac{M}{v+h} \right)^\frac{4}{3} \right] (v+h)^2 \left| \partial U \right|^2,
\end{equation}
where $M \equiv \left( \frac{\kappa^2}{\lambda_\Sigma} \right)^\frac{1}{2}$.

From $K$ and $F$, one may get sectional curvatures $\mathcal{K}_h$ and $\mathcal{K}_\pi$, as defined in \cite{Cohen:2021ucp}:
\begin{equation}
\label{eq:kh}
    \mathcal{K}_h=-\frac{1}{K^2} \left( \frac{F''}{F} - \frac{K' F'}{K F}\right),
\end{equation}

\begin{equation}
\label{eq:kp}
    \mathcal{K}_\pi=-\frac{1}{v^2 F^2} \left( 1 - \frac{v^2 F'}{K^2}\right),
\end{equation}
where the prime denotes a derivative with respect to $h$. Using these forms, the matrix element, $\mathcal{M}$, is written down like in Ref.~\cite{Cohen:2021ucp}  as

\begin{equation}
\label{eq:nh}
     \mathcal{M}_{V_L V_L \to h^n} = -\left( s_{12} - \frac{2 m_h^2}{n+1} \right) \left(\frac{\partial_h}{K(h)}\right)^{n-2} \mathcal{K}_h(h) \bigg|_{h=0},
\end{equation}
where $n=2$ or $3$ for both the $2\to 2$ case and the $2\to 3$ case. The gauge boson scattering process has the matrix element 

\begin{align}
\label{eq:vv}
    \mathcal{M}_{V_L V_L \to V_L V_L} &= \frac{1}{6} T_{ijkl} s_{1234}  \\ \nonumber
    &-\frac{1}{2} \left[ \sigma_{ij} \sigma_{kl}(s_{12}+s_{34})+\sigma_{ik} \sigma_{jl}(s_{13}+s_{24})+\sigma_{il} \sigma_{jk}(s_{14}+s_{23}) \right] \mathcal{K}_\pi(h) \bigg|_{h=0},
\end{align}
where $\sigma_{ij}$ and $T_{ijkl}$ are defined as in Eq.~\eqref{eq:sigma}, and $s_{ij...k}=(p_i+p_j+\cdots p_k)^2$, with all the momenta pointing inwards. For the $V_L V_L \to V_L V_L h$ process, one just replaces $\mathcal{K}_\pi(h)$ with $\left(\frac{\partial_h}{K(h)}\right) \mathcal{K}_\pi(h)$ in Eq.~\eqref{eq:vv}. The potential terms from Eqs.~(\ref{eq:vvpot}, \ref{eq:vvhpot}) still contribute, but as these are not energy-growing, they can safely be neglected, as they do not violate perturbative unitarity in $\mathcal{O}$(TeV). 

As can be seen, the matrix elements have energy dependence, both for the $2\to 2$ and the $2\to 3$ processes. This means that there will be unitary violation for both sets of processes. However, one thing of note is that when all $V_L=Z_L$ in Eq.~\ref{eq:vv},

\begin{subequations}
    \begin{align}
        &s_{12}+s_{34}+s_{13}+s_{24}+s_{14}+s_{23}= 8m_Z^2,& \quad &s_{1234}=0, &\mbox{($2\to 2$ case)}, \\
        &s_{12}+s_{34}+s_{13}+s_{24}+s_{14}+s_{23}= 8m_Z^2 +m_h^2,& \quad &s_{1234}=m_h^2,  &\mbox{($2\to 3$ case)}.
    \end{align}
\end{subequations}
This shows that the $Z_L Z_L \to Z_L Z_L h$ only has a constant modification to $\mathcal{M}$, even for the derivative terms, similar to the potential modification cases. For this reason, the $Z_L Z_L \to Z_L Z_L h$ process is omitted in this analysis.

As before, only the contact term is kept. However, complications arise due to the fact that now, the matrix elements are momentum-dependent, so the phase space integral, $\int \mathrm{dLIPS}_3 \mathcal{M}$, has non-trivial angular and momentum parts, which makes computing $\hat{\mathcal{M}}$ more difficult. Due to this, a numerical approach was taken, where a Monte Carlo integration was done to obtain $\mathrm{dLIPS}_3$. For more details, see App.~\ref{app:dlips}.

\subsection{Results for the derivative terms}

The results are seen in Fig.~\ref{fig:res2}, and the tabulated values in Tables~\ref{table:res22} and  \ref{table:res23} of App.~\ref{app:values}. There are a few things to note. First of all, we show that the $V_LV_L h$ processes are a viable alternative to the $hhh$ processes. The same tendency is seen in the $2\to 2$ processes, where the $V_L V_L$ final states have similar or lower $E_*$ values compared to the $hh$ final state. As mentioned in Sec.~\ref{sec:PUV}, the mass scale of the new particles are lower than any of the calculated $E_*$ values. For example, in the benchmark points, we have $m_f= 452$ GeV ($y=2.6$) and $m_S= 501$ GeV, much lower than the lowest possible $E_*$ value in both cases (3.1 TeV in the $W_L W_L \to Z_L Z_L$ and  $Z_L Z_L \to W_L W_L$ processes in the fermion case with $y=2.6$ and 8.9 TeV in the $W_L W_L \to Z_L Z_Lh$ and  $Z_L Z_L \to W_L W_Lh$ processes in the scalar singlet case). So if these theories are true in nature, the particles will be seen before reaching  the given $E_*$ value.~\footnote{All the loryons searched in Ref.~\cite{Banta:2021dek} have masses of $800$ GeV or less.} As mentioned in Sec.~\ref{subsec:mder}, the potential terms are neglected for the $E_*$ computation. This is because, as seen by comparing Figs.~\ref{fig:res1} and \ref{fig:res2}, the unitarity violation scales for the derivative terms are much lower than those of the potential terms. The potential contribution barely changes the $E_*$ value.

Next, two benchmarks were chosen in the fermion portion to show that the scaling in the $hhh$ case is different from the scaling in the $V_L V_L (h)$ case. This is due to the fact that $\mathcal{K}_h$ and $\mathcal{K}_\pi$ have different dependences on $y$ in the regime of the parameter space chosen. In general, for $\mu=m_f$ and $m=0$, using Eqs.~(\ref{eq:fermion}, \ref{eq:kh}, \ref{eq:kp}),

\begin{equation}
    \mathcal{K}_{h,\Psi}= \frac{3y^4}{8 \pi^2 v^2} \frac{8 \pi^2-y^2}{\left(6 \pi^2-y^2\right)^2},
\end{equation}
\begin{equation}
    \mathcal{K}_{\pi, \Psi}=\frac{y^2}{8 \pi^2 v^2} \frac{16 \pi^2 - 3 y^2}{6 \pi^2-y^2}.
\end{equation}
In the part which is rational in $y$, given the choices in parameters in this paper, the constant terms dominate over the coupling terms, so roughly speaking, $\mathcal{K}_h\propto y^4$ and $\mathcal{K}_\pi \propto y^2$. Then, from Eqs.~(\ref{eq:nh}, \ref{eq:vv}, \ref{eq:Mavg}), it can be shown that for the $2\to 2$ processes, the energy growing behavior is proportional to $|\mathcal{K}_a|^2E^4$, $a=h,\pi$. This would mean for this case if the phase space factor were to be neglected,

\begin{equation}
    \frac{E_i}{E_j} \simeq \sqrt{\frac{|\mathcal{K}_a^j|}{|\mathcal{K}_a^i|}},
\end{equation}
where $i$ and $j$ label two different parameter choices. This means that for the $2\to 2$ case,
\begin{subequations}
    \begin{align}\frac{E_i}{E_j}\simeq\frac{y_j^2}{y_i^2}, \quad \quad\mbox{($hh$ final state),} \end{align}
\begin{align}\frac{E_i}{E_j}\simeq\frac{y_j}{y_i}, \quad \quad \mbox{($V_L V_L$ final state).} \end{align}
\end{subequations}
This shows why the parametric dependence changes from the two processes in the regime where $y^2< 6 \pi^2$. In the case of the 3-body final state, the energy growing behavior is $|\mathcal{K}_a|^2E^6$, as $\mbox{dLIPS}_3 \propto E^2$, and the derivative of $\mathcal{K}_a$ still adds terms where the constants dominate. This means for the $2\to 3$ case, when $y^2< 6 \pi^2$,

\begin{subequations}
    \begin{align}\frac{E_i}{E_j}\simeq \left(\frac{y_j}{y_i} \right)^\frac{4}{3}, \quad \quad\mbox{($hhh$ final state)}, \end{align}
\begin{align} \frac{E_i}{E_j}\simeq \left(\frac{y_j}{y_i} \right)^\frac{2}{3}, \quad \quad \mbox{($V_L V_L h$ final state)}. \end{align}
\end{subequations}
This same analysis may be done for the singlet and doublet scalar cases using Eqs.~(\ref{eq:kh}, \ref{eq:kp}, \ref{eq:scalar}, \ref{eq:doublet}). However, for the singlet scalar case, one thing to note is that for $m=0$, $K(h)$ is actually a constant; this leads to a vanishing $\mathcal{K}_h$. This is why there are no low $E_*$ values for the Higgs final state processes. For the $V_L V_L$ and $V_L V_L h$ processes, a similar derivation as before may be done to get that, for $\lambda < 96 \pi^2$,
\begin{subequations}
    \begin{align}\frac{E_i}{E_j}\simeq \sqrt{\frac{\lambda_j}{\lambda_i}},  \quad \quad\mbox{($V_L V_L$ final state),} \end{align}
\begin{align} \frac{E_i}{E_j}\simeq \left(\frac{\lambda_j}{\lambda_i} \right)^\frac{1}{3}, \quad \quad \mbox{($V_L V_L h$ final state).} \end{align}
\end{subequations}
Then, for the SU(2) doublet, if $\kappa^2 < \lambda_\Sigma v^2$, the two cases have similar behavior, and  
\begin{subequations}
    \begin{align}\frac{E_i}{E_j}\simeq \left(\frac{M_j}{M_i}\right)^\frac{2}{3},  \quad \quad\mbox{(2-body final state),} \end{align}
\begin{align}\frac{E_i}{E_j}\simeq \left(\frac{M_j}{M_i}\right)^\frac{4}{9},  \quad \quad\mbox{(3-body final state).} \end{align}
\end{subequations}
The point $\lambda=8.3$ for the scalar and $y=2.6$ for the fermion are the maximum possible coupling values for no bare mass given in the results of \cite{Banta:2021dek}. This was done to compare the lowest unitarity violation scales for the given parameter space, to show that the $V_LV_L(h)$ processes are compared to the $h^n $ processes. Using these two values and their given energies in Fig.~\ref{fig:res2}, any other $E_*$ value in the allowed regimes with no bare mass can be found using the above equations.

\section{Discussion and Conclusion}

Finding the shape of the Higgs potential is one of the most important current goals in physics. While the di-Higgs boson search is a very promising probe of the Higgs cubic coupling, looking at gauge boson/Higgs scattering will allow us to probe the Higgs potential independently as well. Hence, high energy electroweak boson scattering processes are essential and complementary to the di-Higgs boson searches. However, experimental programs at high multiplicities are not yet established.

 From this point of view, we ask, ``Is the process with multiple gauge bosons in the final state just as good as the one with multiple Higgs bosons to examine modifications of the Higgs sector at high-energy colliders?'' Answering this question was the main goal of this work, and it is done in two ways. Firstly, we evaluate the energy scales of unitarity violation for the Higgs cubic coupling shift as well as for multiple different Higgs potential modifications that fall under the HEFT. We use the equivalence theorem and compute scattering processes at the tree level. Our results in Fig.~\ref{fig:res1} suggest that the $V_L V_L\to V_L V_L h$ processes have similar unitarity violation scales to the $V_L V_L\to h h h$ process. Secondly, we also examine the $2\to 2$ scattering because the energy growth would appear in this case if HEFT derivative interactions exist. Such derivative terms are common for many HEFTs, which are obtained from known UV models. We show, in Fig.~\ref{fig:res2}, that the unitarity violation scales of the $V_L V_L \to V_L V_L$ processes are similar or even lower than that of the $V_L V_L \to h h$ process. We find the same tendency between the $V_L V_L \to V_L V_L h$ processes and the $V_L V_L \to hhh$ processes. Furthermore, we address the experimental advantage in the final states with multiple gauge bosons in Table~\ref{table:1}.  

Let us clarify our method regarding $E_*$, the scale of perturbative unitarity violation in the scattering. We use the value of $E_*$ to compare different processes in each modification of the Higgs sector, mainly between $hh(h)$ and $V_LV_L(h)$ final states. 
We checked in several cases our $E_*$ values are consistent with Ref.~\cite{Abu-Ajamieh:2020yqi, Falkowski:2019tft, Chang:2019vez, Cohen:2021ucp}. 
Note that reaching the given $E_*$ value was not the main focus of this paper. The mass of new particles can be much lower than the $E_*$ value, and the values of $E_*$ depend on the size of couplings. However, the relative importance of different processes is robust.

Given that prospective modification of the Higgs sector is unknown, experiments need to set up observables that cover a wide class of models and EFTs and are simultaneously more feasible. We emphasize that this paper's results will help pick up good observables, such as $V_L V_L\to V_L V_L h$,  that can be sensitive to not only the cubic coupling but also other possible modifications.   
For the $2\to2$ processes, we even show that all the considered HEFT interactions are probed better in the $V_L V_L\to V_L V_L$ processes. This result enhances the impact of the current experimental program at the LHC.

For realistic sensitivities, one needs to perform the background studies to see whether or not the $V_L V_L h$ final state has an advantage over the $hhh$  in the various collider setups, including the HL-LHC, FCC, and muon collider. However, they are beyond the scope of this paper. If one confirms that the background level for both processes is similar, the promising benchmark $2\to3$ process for exploring generic modifications of the Higgs potential is $V_L V_L\to V_L V_L h$ because the $V_L V_L h$ final state has higher branching ratios in the cleaner decay modes. For the $V_LV_L\to V_LV_L$ processes, it is necessary to look at the high energy part of the di-Higgs process from the vector-boson fusion, while the CMS experiment tested the $V_LV_L\to V_LV_L$ process~\cite{CMS:2023rcv}. 
Beyond 3-body final states, an interesting process is $V_L V_L\to  V_L V_L V_L V_L$ as discussed in Ref.~\cite{Chang:2019vez}, since this process has much smaller $E_*$ values in the case of cubic coupling modification, which could be true for other modifications. However, this has the complications of being a 4-body final state, leading to many diagrams (of $\mathcal{O}(10^5)$) and tiny cross sections. The practical sensitivity studies need more computational efforts.

%%%%%%%%%%%%%%%%%%%%%%%%%%%%%%%%%%%%%%%%%%%%%%%%%%%%%%%%%%%%%%%%%%%%%%%%%%%%
%%%%%%%%%%%%%%%%%%%%%%%%%%%%%%%%%%%%%%%%%%%%%%%%%%%%%%%%%%%%%%%%%%%%%%%%%%%%
\section{Acknowledgements}
We thank Maria Mazza, Takemichi Okui, Xiaochuan Lu, and Yoshihiro Shigekami for the useful discussion. This work was supported by, in part, the US Department of Energy grant DE-SC0010102. KT is also supported by JSPS KAKENHI 21H01086 and FSU Summer Research Support award Program.
%%%%%%%%%%%%%%%%%%%%%%%%%%%%%%%%%%%%%%%%%%%%%%%%%%%%%%%%%%%%%%%%%%%%%%%%%%%%

\appendix
\section{How many Higgs bosons can be replaced by longitudinal gauge bosons?}\label{app:replacing}

In general, the number of Higgs bosons that can be replaced by the longitudinal gauge bosons in some $V_L V_L \to h^n$ process depends on the type of new interactions. Firstly, all the NG bosons must be added in pairs as 
\begin{subequations}
    \begin{align}
        &H^\dag H \supset 2 v h + h^2+ G^2  \quad \mbox{(Potential terms),}\\
        &|\partial H|^2 = \frac{1}{2}( \partial h)^2 + \frac{1}{2}(\partial G)^2 \quad \mbox{(Derivative terms),}
    \end{align}
\end{subequations}
so if there is an odd number of Higgs bosons, one of them must remain. 

Let us consider a shift in some power of the Higgs potential, $h^\ell$. This, to the lowest order in the NG bosons, can be written as
\begin{align}
\label{eq:Gh}
    h^\ell \to \left(\sqrt{2 H^\dag H} - v \right)^\ell = \left(h+ \frac{G^2}{2(v+h)} + \cdots\right)^\ell.
\end{align}
Here, by expanding, we get $G^{2\ell}$ as the lowest contact interaction of only $G$, and every term with a lower power of $G$ has to be multiplied by some number of $h$'s. This shows that if an $h^\ell$ term is added to the potential, the $G^{2(\ell- k)}h^k$, $k<\ell$ contact interaction can not have any Higgs bosons replaced. This is why, for example, the shift of Higgs cubic coupling gives no $G^4$ contact interaction, while $G^2 h^2$ interaction exists but does not cause the unitarity violation. Another example is that the shift of Higgs quartic coupling brings a $G^2 h^3$ interaction leading to the $V_LV_L\to hhh$ unitarity violation, but the $G^4 h$ interaction is absent thus no unitarity violation in $V_LV_L\to V_L V_Lh$.

 From Eq.~\eqref{eq:Gh}, one can find a $G^{2k}h^{2(\ell -k)}$ interaction by expanding the $\frac{1}{v+h}$ part, leading to the $V_L V_L \to V_L^{2(k-1)} h^{2 (\ell - k)}$ process for $k<\ell$. Here, all the $\ell- k$ Higgs boson pairs of this process can be replaced by the gauge bosons because the $G^{2\ell}$ contact interaction exists. More specifically, for $k=1$, in the $V_L V_L \to h^{2 (\ell - 1)}$ process,  it is possible to replace all Higgs boson pairs. 

If there is an odd number of Higgs bosons, a $2\to(2\ell-3)$ process can also have all Higgs boson pairs replaced as well, leaving just one Higgs boson. This is why in the $G^2h^3$ case for a cubic coupling shift, two Higgs bosons can still be replaced. In general, for a $G^{2 k} h^{2(\ell-k)-1}$ contact interaction for $k<\ell$, $\ell-k-1$ Higgs boson pairs may be replaced by NG boson pairs due to the presence of $G^{2\ell-2}h$ contact interaction.

In the case of some other potentials, such as those derived from UV models, the situation changes because they are typically a function of $H^\dag H$, not $\sqrt{2 H^\dag H} - v$. This means that in a generic case, there is always a constant term, such as $v^2$, that may be multiplied by the $G^2$ terms after expanding the potential. For example, consider Eq.~\eqref{eq:log},

\begin{align}
  \left( H^\dag H \right)^2 \ln\left( H^\dag H \right) \supset & \frac{v^4}{2} \ln\left( 1+ \frac{h}{v} \right) + \frac{v^4}{4}\ln\left(1+\frac{G^2}{\left(v+h \right)^2}\right) \\ \supset & v^2\frac{G^2}{4} \sum_{n=0}^\infty \left( n+1 \right) \left(-\frac{h}{v}\right)^n + 
\frac{G^4}{8} \sum_{n=0}^\infty \frac{\left( n+1 \right) \left( n+2 \right) \left( n+3 \right)}{6} \left(-\frac{h}{v}\right)^n + \cdots \nonumber
\end{align}
Here, the second term gives an infinite number of $G^2h^n$ terms, but because every higher order $G^{2}$ term has a series expansion in $h$ starting with a constant, any number of (even) Higgs bosons may be replaced by NG bosons. This is also true for $\left(H^\dag H\right)^{\frac{2}{3}}$ or $\sqrt{H^\dag H}$ as well, as most UV models do not give an EFT with a function like $\sqrt{2 H^\dag H} - v$.

Another remark is that going the ``other way'', i.e., replacing NG boson pairs with Higgs boson pairs is also not always guaranteed to have energy-growing behavior. For example, consider the HEFT interactions involving derivatives from the singlet scalar model, which is discussed in Sec.~\ref{sec:HEFTderivative}. We show that the $V_LV_L \to hh$ does not violate unitarity in some cases, while the $V_LV_L \to V_LV_L$ case does. This is understood by the different curvatures $\mathcal{K}_{h,S}$ and $\mathcal{K}_{\pi,S}$,
\begin{subequations}
    \begin{align}
    \mathcal{K}_{h,S} &= \frac{1}{48 \pi^2}\frac{\lambda^2 m^2}{(2m^2 + \lambda (1+ \frac{\lambda}{96 \pi^2}) (v+h)^2)^2}, \\ 
     \mathcal{K}_{\pi,S} &= \frac{1}{96 \pi^2}\frac{\lambda^2}{2m^2 + \lambda (1+ \frac{\lambda}{96 \pi^2}) (v+h)^2}.
    \end{align}
\end{subequations}
which are given in~Ref.~\cite{Cohen:2021ucp}. 
If the bare mass $m$ is non-zero, the unitarity violation appears in the $V_LV_L \to h^n$ processes. 
However, as $m$ gets smaller, $\mathcal{K}_{h,S}$ decreases making $V_LV_L \to h^n$ processes smaller, while the $V_LV_L \to V_LV_L(h)$ process shows a larger unitarity violation.
In the limit of $m=0$, the unitarity violation exists in $V_LV_L \to V_LV_L(h)$  but not in $V_LV_L \to hh(h)$. 
This shows the reverse case of the $h^\ell$ coupling shift
where the $G^4(h)$ is the leading contact interaction which has $2 \to 2(3)$ unitarity violation. And, since $ \left(\frac{\partial_h}{K(h)} \right)^n \mathcal{K}_{h,S} =0$ for all $n$ in the massless case, all $V_LV_L \to h^n$ processes will lack the stronger unitarity violation from derivative terms in this parameter choice. Note that the weaker unitarity violation from UV potential modifications still exists for higher multiplicities of $V_LV_L \to h^n$, but they will give much higher $E_*$ values compared to new derivative terms, as demonstrated in Secs.~\ref{sec:Higgs-potential} and \ref{sec:HEFTderivative}. 
%%%%%%%%%%%%%%%%%%%%%%%%%%%%%%%%%%%%%%%%%%%%%%%%%%%%

\section{Average Matrix Element and Unitarity Violation}\label{app:mhat}
The following discussion is taken from App.~A in \cite{Chang:2019vez} and App. B in \cite{Cohen:2021ucp}.
To get bounds from perturbative unitarity, the $S$-matrix and the optical theorem are used. Firstly, writing the $S$ matrix,~\footnote{This is, again, following convention of \cite{Cohen:2021ucp}, so $T$ is negative of that in ex. Peskin and Schroeder.}

\begin{align}
    S=1-i T,
\end{align}
where $T$ denotes the interactions of the particles. Then consider the quantum state $\ket{ p, \alpha} $, where $p$ is the momentum, and $\alpha$ is other quantum mechanical information assumed to be discrete (for example, relative angular momentum or particle type). The normalization condition is

\begin{align}
\label{eq:norm}
    \braket{q, \beta | p, \alpha} = (2 \pi)^4 \delta^4(p-q) \delta_{\alpha \beta}.
\end{align}
and the matrix element is defined as

\begin{align}
 & \braket{q, \beta |S| p, \alpha} = (2 \pi)^4 \delta^4(p-q) S_{\alpha \beta}, \\ \label{eq:Mavgdef}
  & \braket{q, \beta |T| p, \alpha} = (2 \pi)^4 \delta^4(p-q) \hat{\mathcal{M}}_{\alpha \beta},
\end{align}
to factor out the delta function. We obtain
\begin{align}
    S_{\alpha \beta} = \delta_{\alpha \beta} - i \hat{\mathcal{M}}_{\alpha \beta}.
\end{align}
By unitarity arguments, $|S_{\alpha \beta}|^2\leq1$. Setting $\alpha \neq \beta$ gives 

\begin{align}
    |\hat{\mathcal{M}}_{\alpha \beta}|^2 \leq 1. 
\end{align}
Next, consider $\alpha= \beta$. We apply the optical theorem, 

\begin{align}
   & 1 = \sum_\beta |S_{\alpha \beta}|^2= 1 + 2 \Im \hat{\mathcal{M}}_{\alpha \alpha} + \sum_\beta |\hat{\mathcal{M}}_{\alpha \beta}|^2   \\ 
    &\to  -2 \Im \hat{\mathcal{M}}_{\alpha \alpha} = \sum_\beta |\hat{\mathcal{M}}_{\alpha \beta}|^2 \geq |\hat{\mathcal{M}}_{\alpha \alpha}|^2= | \Re \hat{\mathcal{M}}_{\alpha \alpha}|^2 + |\Im \hat{\mathcal{M}}_{\alpha \alpha}|^2  \\ \label{eq:puv}
     &\to 1 \geq | \Re \hat{\mathcal{M}}_{\alpha \alpha}|^2 + |\Im \hat{\mathcal{M}}_{\alpha \alpha}+1|^2.
\end{align}

From Eq.~\eqref{eq:puv}, we get that $|\Re \hat{\mathcal{M}}_{\alpha \alpha}| \leq 1$, and $0 \geq \Im \hat{\mathcal{M}}_{\alpha \alpha} \geq -2$. As the analysis is a tree-level one, we may omit the imaginary part restriction and say that 
\begin{align}
    |\hat{\mathcal{M}}_{\alpha \beta}| \leq 1, \quad \quad \forall \alpha, \beta \quad \quad \mbox{(tree level),}
\end{align}
Now, we have a unitarity condition for any particle state $\alpha$ and $\beta$. Next, for the normalization factor to the canonically normalized matrix element, $\mathcal{M}$, we use Eqs.~(\ref{eq:norm}, \ref{eq:Mavgdef}), and for simplicity, use a set of scalar states $1, 2, \cdots r$,

\begin{align}
\label{eq:state}
    \ket{p, k_1, k_2, \cdots , k_r} =  A_{k_1, k_2, \cdots, k_r}\int \mathrm{d}^4 x e^{i p  \cdot x} \prod_{i=1}^r \left[\phi_i(x) \right]^{k_i} \ket{0},
\end{align}
where $\phi_i(x)$ is the field with the creation operator for the $i$th particle. By braketing the $T$-matrix with $\bra{q, m_1, m_2, \cdots , m_r}$ and $\ket{p, n_1, n_2, \cdots , n_r}$ and applying Eq.~\eqref{eq:Mavgdef}, we can relate the canonically normalized matrix element $\mathcal{M}$ to the unitarity bounded matrix element as

\begin{align}
\label{eq:mhatnoconst}
    \hat{\mathcal{M}}_{\alpha \beta} = A^*_{n_1, n_2, \cdots, n_r} A_{m_1, m_2, \cdots, m_r} \int \mathrm{dLIPS}_m \int \mathrm{dLIPS}_n \mathcal{M}_{\alpha \beta},
\end{align}
where $m_i$ and $n_i$ denote identical initial and final particles, respectively. Note that $\hat{\mathcal{M}}$ only picks out $s$-wave states, as they have the simplest unitarity behavior (an example of more complicated behavior would be the $t$ and $u$ channels each giving a constant term in the $ZZ\to hh$ cross section despite the matrix element being constant as well). All that is left now is to find the normalization factor $A_{k_1, k_2, \cdots, k_r}$, and this is done by braketing the states in \eqref{eq:state} and applying Eq.~\eqref{eq:norm}. It can be shown that 
\begin{align}
    \frac{1}{|A_{k_1, k_2, \cdots, k_r}|^2} = \prod_{i=1}^r k_i! \int \mathrm{dLIPS}_k,
\end{align}
where $\sum_{i=1}^r k_i = k$, and dLIPS$_k$ is defined in Eq.~\eqref{eq:LIPSn}. Then plugging this result into Eq.~\eqref{eq:mhatnoconst} finally gives Eq.~\eqref{eq:Mavg}, where the indices are dropped for brevity:

\begin{equation}
\hat{\mathcal{M}} = \left( \frac{1}{ \prod_i m_i! \prod_i n_i! \int \mathrm{dLIPS}_m \int \mathrm{dLIPS}_n}\right)^\frac{1}{2} \int \mathrm{dLIPS}_m \int \mathrm{dLIPS}_n \mathcal{M}.
\end{equation}

\section{Cross section growth for higher multiplicity final states}\label{app:x-sec}

We use the phase-space averaged matrix element, $\hat{\mathcal{M}}$, to quantify the perturbative unitarity violation of electroweak boson scattering. In this appendix, we relate $\hat{\mathcal{M}}$ to the cross section. We show that a phase-space averaged matrix element growing faster in energy, which is typically at higher multiplicities, also leads to a fast growth in the cross section.

For simplicity, we assume the massless limits, an initial state of $W_L W_L$, and a constant matrix element $\cal M$. From Eq.~\eqref{eq:Mavg}, the averaged matrix element of the $2\to n$ process is given by 

\begin{subequations}
\begin{align}
    \hat{\mathcal{M}} = & \left( \frac{1}{ 8 \pi \Pi_i n_i! \int \mathrm{dLIPS}_n}\right)^\frac{1}{2}  \int \mathrm{dLIPS}_n \mathcal{M} \ \to
    \\
    &|\hat{\mathcal{M}}|^2 = |{\mathcal{M}}|^2
    \left( \frac{ \int \mathrm{dLIPS}_n}{ 8 \pi \Pi_i n_i!}\right)  , 
\end{align}
\end{subequations}
where we use $\mathcal{M}$ being constant in the second line. 
Since the cross section is proportional to
\begin{equation}
\sigma \propto \frac{1}{E_{\rm cm}^2} \int \mathrm{dLIPS_n} \left| \mathcal{M}\right|^2 \ , 
\end{equation}
 the relation with  $\hat{\mathcal{M}}$ is 
\begin{equation}
    \sigma \propto \frac{8 \pi \Pi_i n_i!}{E_{\rm cm}^2} |  \hat{\mathcal{M}}|^2.
\end{equation}
The explicit energy dependence of the cross section is seen using Eq.~\eqref{eq:LIPSn},
\begin{equation}
    \sigma \propto \frac{\left| \mathcal{M}\right|^2}{ \left(n-1 \right)! \left(n-2 \right)!} \left( \frac{E_{\rm cm}}{4 \pi} \right)^{2(n-3)}.
\end{equation}
This shows that for a constant matrix element, the energy dependence of the cross section gets higher powers for the larger multiplicities. For the HEFT interactions, most powers of $E_{\rm cm}$ are compensated by the Higgs VEV, such as $(E_{\rm cm}/v)$. Therefore, in the case of HEFT, a process with higher multiplicities tends to have a lower energy scale of unitarity violation. %
This statement is supported by Ref.~\cite{Chang:2019vez}, as it recommends looking at the $V_L V_L \to V_L V_L V_L V_L$ process at the collider. Hence, if the goal is to have an $E_*$ value which can be accessible to colliders, the $V_L V_L \to V_L V_L V_L V_L$ process would be a better candidate to search for rather than one of the 3-body final state processes.

\section{Evaluating the phase space integral}
\label{app:dlips}
Here, we compute

\begin{equation}
    \mathrm{dLIPS}_n = \prod_{i=1}^n \frac{\mathrm{d}^3p_i}{(2 \pi)^3 2 E_i}
 (2 \pi)^4 \delta^4\left(\sum_i p_i\right).
 \label{eq:dlips}
 \end{equation}
 In this paper, we need to evaluate the dLIPS$_n$ integral for the $n=2$ and 3 cases. We can perform the integral with $n=2$ analytically, and the result is 
 \begin{equation}
 \label{eq:dlips2}
     \int  \mathrm{dLIPS}_2 = \frac{F}{8 \pi}, 
 \end{equation}
where $F=\sqrt{\left(1-\frac{m_1^2}{s}-\frac{m_2^2}{s} \right)^2-4 \frac{m_1^2 m_2^2}{s^2}}$ for the masses of $m_{1,2}$. This simplifies to $F=\sqrt{1-\frac{4 m^2}{s}}$ for the same masses $m_1=m_2=m$. 

In general, for $2\to 2$ scattering, all the kinematics can be expressed in terms of $s$, $t$, and $u$, the Mandelstam variables.  The $s$ variable is independent of the integral, keeping the results the same. Regarding $t$ and $u$ dependence, the contact term matrix element, $\mathcal{M}$, will have terms proportional to $\cos \theta$, where $\theta$ is the final state momenta's angle.  Those terms vanish due to the fact that $\int \mathrm{dLIPS}_2 \mathcal{M} \propto \int_{-1}^1 \cos \theta \mathrm{d}\cos \theta$. This leaves only constant terms, so one can use Eq.~\eqref{eq:dlips2}.

For $n=3$, we use the integration method similar to the one used in Ref.~\cite{Girmohanta:2023tdr}. We chose to keep the mass despite using the equivalence theorem, as it did not add much extra complexity to the Monte-Carlo integration.
Three of the integrals, for example, the Higgs boson momentum integrals in $V_LV_L h$, can be removed by the delta functions, leading to
\begin{equation}
   \mathrm{dLIPS}_3= \frac{1}{2 E_3} \frac{\mathrm{d}^3p_1}{(2 \pi)^3 2 E_1} \frac{\mathrm{d}^3p_2}{(2 \pi)^3 2 E_2} 2 \pi \delta(E_{\rm cm} - E_1-E_2-E_3),
\end{equation}
where $E_{1,2}^2=p_{1,2}^2+m_{1,2}^2$, $E_3= \sqrt{p_1^2+p_2^2 +2 p_1 p_2 f+m_h^2}$, $m_h$ is the Higgs boson mass, and $p_1$ and $p_2$ are the magnitudes of the three-momenta of the two $V_L$. 
And, $f$ represents
\begin{equation}
\label{eq:f}
    f \equiv f(\theta, \phi)= \cos \theta_1 \cos \theta_2 + \sin \theta_1 \sin \theta_2 \sin \Delta \phi= x_1 x_2+\sqrt{1-x_1^2} \sqrt{1-x_2^2} \sin \Delta \phi,
\end{equation}
where  $\theta_1$ and $\theta_2$ are the angles of $\overrightarrow{p_1}$ and $\overrightarrow{p_2}$ respectively, $x_i =\cos \theta_i$ and $\Delta \phi =\phi_1-\phi_2$ is the difference of the azimuthal angles of $\overrightarrow{p_1}$ and $\overrightarrow{p_2}$. The range of $f$ is $[-1,1]$. A change of variables is made such that  $\Delta \phi \to \phi_2$ as well, and then one of the $\phi$ integrals is trivial giving a factor of $2 \pi$. So, the 3-body integral becomes
\begin{equation}
    \int \mathrm{dLIPS}_3 = \int \frac{p_1^2\mathrm{d}p_1 p_2^2 \mathrm{d}p_2 \mathrm{d}x_1 \mathrm{d}x_2 \mathrm{d}\phi_2}{8 (2 \pi)^4 E_1 E_2 E_3 } \delta(E_{\rm cm}-E_1-E_2-E_3),
\end{equation}
where $E_{1,2}$ is a function of $p_{1,2}$ and $m_{1,2}$, and $E_3$ is a function of those as well as $x_1$, $x_2$, and $\phi$, as defined above Eq.~\eqref{eq:f}. Now, the common delta function property is used,
\begin{equation}
    \delta(g(x)) =\sum_i \frac{\delta(x-x_i)}{|g'(x_i)|},
\end{equation}
where $g(x_i)=0$. This can be used to remove one of the $p_i$'s (for the case with identical $V_L$, this choice is irrelevant; otherwise, assume the heavier particle, $p_2$, is chosen).  In this case,

\begin{figure}[!tbp]
\label{fig:ps}
\centering
\includegraphics[width=300 pt]{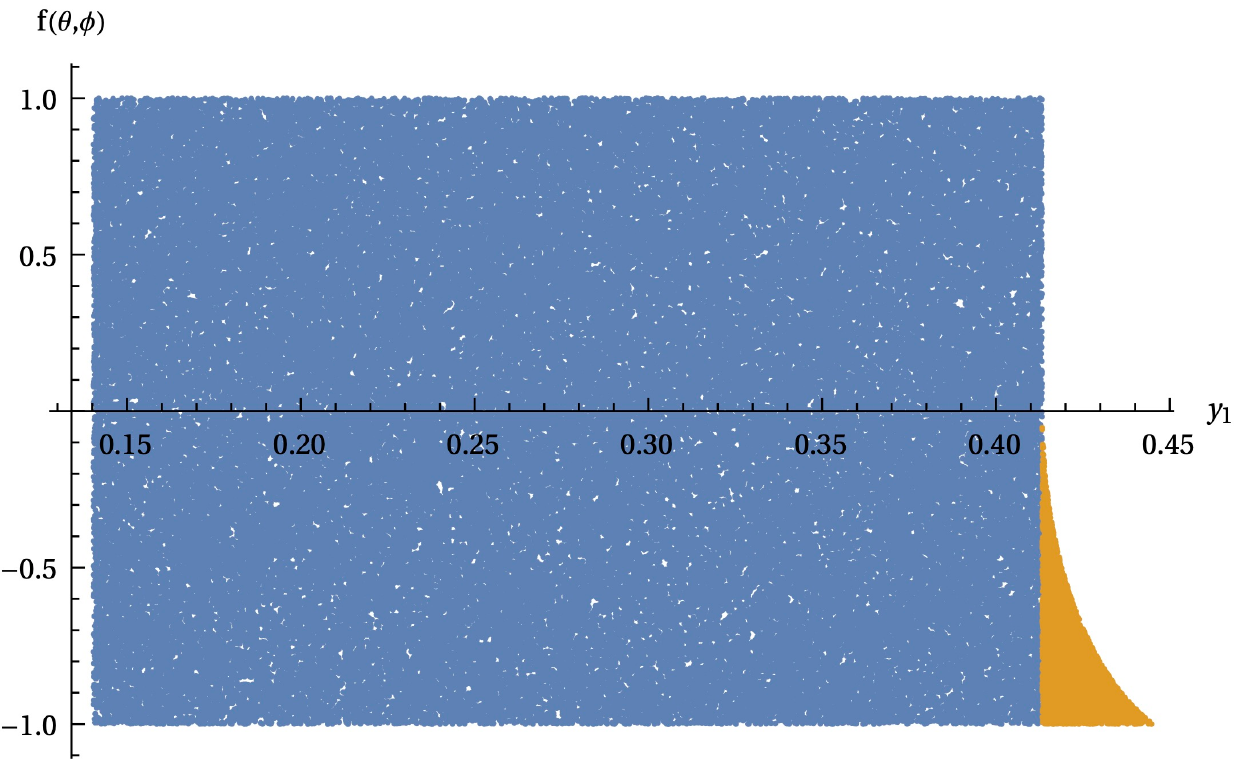}
\caption{Example of the Monte-Carlo 3-body phase space integration region for the $W_L W_L \to W_L W_L h$ case using $E_{\rm cm}=2(2 m_W+m_h)$. Here, $y_1=\frac{E_1}{E_{\rm cm}}$, and $f(\theta, \phi)$ is defined in Eq.~\eqref{eq:f}. For the solution $p_2^+$, both the orange and blue regions are integrated over, as $p_2^+$ is real and non-negative in both regions.  For $p_2^-$, only the orange region has real and non-negative values.  The $y_1 E_{\rm cm} < m_W$ region is also removed due to the fact that $E_1\geq m_W$.}.
\end{figure}

\begin{align}
    &g(p_2)=E_{\rm cm}-E_1-E_2-E_3 \\
    &g'(p_2) = \frac{p_2}{E_2}+\frac{p_2+f p_1}{E_3}= \frac{(E_{\rm cm}-E_1) p_2 + E_2 f p_1}{E_2 E_3},
\end{align}
where $E_{\rm cm}=E_1+E_2+E_3$ was used to replace $E_2+E_3$ in the last step, and $E_2$ and $E_3$ are functions of $p_2$ as shown earlier. Using the first condition, one can solve for $p_2$.  Next, using the condition that $p_2$ is real and non-negative, the boundaries for $p_1$ and $f$ are found, and $p_2$ is integrated out. 
It turns out that there are two viable solutions for $p_2$. For example, in the case of $m_1=m_2=m$, the solutions are
\begin{equation}
    p_2^\pm= \frac{-B\pm \sqrt{B^2-4 A C}}{2 A},
\end{equation}
and
\begin{align}
   & A=E_{\rm cm}^2+2 E_{\rm cm} E_1+E_1^2(1-f^2)+m^2 f^2\ ,  \\
   &  B = p_1 f^2 \left(E_{\rm cm}^2- 2 E_{\rm cm} E_1-m_h^2+2m^2 \right),  \\
   &  C = -\left(E_{\rm cm}^2- 2 E_{\rm cm} E_1-m_h^2\right)^2+4\left(p_1^2+m_h^2 \right) m^2.
\end{align}

Both solutions are real and non-negative in different regions of the phase space.  This can be seen in Fig.~\ref{fig:ps}, where different regions are integrated based on which $p_2$ is used. Both the blue and orange regions are integrated for $p_2^+$, and only the orange region is integrated for $p_2^-$. The other regions either have a complex $p_2$ or have $p_2<0$. 

After finding the bounds, a change of variables is done, specifically using the energies instead of the momenta, as $E_1 \mathrm{d} E_1 = p_1 \mathrm{d} p_1$, and then $y_i=\frac{E_i}{E_{\rm cm}}$, and $\tilde{p}_i= \frac{p_i}{E_{\rm cm}}$. Also, we check the consistency of the massive phase space as a function of $\frac{\sum_i m_i}{E_{\rm cm}}$ with \cite{Cohen:2021ucp}. Here, $\sum_i m_i$ is the sum of all the final state masses.

Finally, it is obtained that
\begin{equation}
    \mathrm{dLIPS}_3 = \frac{E_{\rm cm}^2}{128 \pi^4} \frac{\tilde{p_1} \tilde{p_2}}{1-y_1 + \frac{\tilde{p_1}}{\tilde{p_2}} y_2 f} \mathrm{d}y_1 \mathrm{d}x_1 \mathrm{d}x_2 \mathrm{d}\phi_2,
\end{equation}
where either the $p_2^+$ or $p_2^-$ solution is taken based on the region of the phase space. 

After this, a Monte-Carlo integration code was written in Mathematica. The size of random sampling is $10^5$.  An example of the phase space integral region is shown in Fig.~\ref{fig:ps}.  We scan multiple $E_{\rm cm}$ values and interpolate them to obtain the final $ \int \mathrm{dLIPS}_3$ as a function of $E_{\rm cm}$. 

If the matrix element $\mathcal{M}$ has a dependence on any of the variables being integrated, such as in all $V_L V_L \to V_L V_L h$ cases in Sec.~\ref{sec:HEFTderivative}, those parts are included in the numerical integration, i.e., the integral being evaluated is
\begin{equation}
     \int \mathcal{M}\mathrm{dLIPS}_3 = \frac{E_{\rm cm}^2}{128 \pi^4} \int \frac{\tilde{p_1} \tilde{p_2}}{1-y_1 + \frac{\tilde{p_1}}{\tilde{p_2}} y_2 f} \mathrm{d}y_1 \mathrm{d}x_1 \mathrm{d}x_2 \mathrm{d}\phi_2 \mathcal{M}.
\end{equation}

\section{Tabulated values of the results}\label{app:values}

In this section, we include the tabulated version of the results in Figs.~(\ref{fig:res1}, \ref{fig:res2}).

\begin{table}[htb]
\centering
    \begin{tabular}{||c|c|c|c|c||}
    \hline
    Process &  $ \frac{y^2}{(4 \pi^2)}(H^\dag H)^2 \ln(H^\dag H)$ & $ \kappa_\frac{2}{3}( H^\dag H)^\frac{2}{3}$ & $  \frac{m_h^2}{2v} \delta\left(\sqrt{2 H^\dag H} -v \right)^3$ & $\kappa_\frac{1}{2}(H^\dag H)^\frac{1}{2}$\\ 
    \hline 
    $W_L W_L \to hh h$ & 2.7 & 1.7 & 0.12 & 0.68  \\
    \hline
    $Z_L Z_L \to Z_L Z_L h$ & 0.72 & 0.77 & 0.094 & 0.37 \\
    \hline
    $W_L Z_L \to W_L Z_L h$ & 1.1 & 1.2 & 0.14 & 0.55 \\
    \hline
     $Z_L Z_L \to W_L W_L h$ & 1.6 & 1.6 & 0.20 & 0.79 \\
    \hline
     $W_L W_L \to Z_L Z_L h$ & 1.6 & 1.6 & 0.20 & 0.79  \\
    \hline  
    $W_L W_L \to W_L W_L h$ & 0.55 & 0.58& 0.071  & 0.28 \\
    \hline
    \end{tabular}
    \caption{Energy scales of unitarity violation for the bottom-up potential terms that have been added to the SM Lagrangian given in $10^3$~TeV. This corresponds to Fig.~\ref{fig:res1}}
    \label{table:res1}
\end{table}

\begin{table}[htb]
\centering
    \begin{tabular}{||c|c|c|c|c||}
    \hline
    Process &  Fermion $(y=1.4)$ & Fermion $(y=2.6)$ & Scalar singlet & Scalar doublet \\ 
    \hline 
    $W_L W_L \to hh $ & 25 & 6.9 & - & 9.7  \\
    \hline
     $W_L W_L \to W_L W_L$ & 6.8 & 3.7 & 18 & 9.0 \\
    \hline
     $W_L W_L \to Z_L Z_L$ & 5.7 & 3.1 & 15 & 7.6 \\
    \hline
     $Z_L Z_L \to W_L W_L$ & 5.7 & 3.1 & 15 & 7.6 \\
    \hline  
    $W_L Z_L \to W_L Z_L$& 6.8 & 3.7 & 19  & 9.0 \\
    \hline
    \end{tabular}
    \caption{Energy scales of unitarity violation for the chosen HEFTs involving derivatives that modify the SM Lagrangian given in TeV for the $2\to 2$ processes. For the scalar singlet, $\lambda=8.3$, for the scalar doublet, $M= 45$ GeV, where $M^2= \frac{\kappa^2}{\lambda_\Sigma}$. The $hh$ final state has no unitarity violation scale from the derivative terms when the bare mass is set to zero. This corresponds to the left plot in Fig.~\ref{fig:res1}}
    \label{table:res22}
\end{table}

\begin{table}[htb]
\centering
    \begin{tabular}{||c|c|c|c|c||}
    \hline
    Process &  Fermion $(y=1.4)$ & Fermion $(y=2.6)$ & Scalar singlet & Scalar doublet \\ 
    \hline 
    $W_L W_L \to hhh $ & 14 & 7.2 & - & 6.1  \\
    \hline
     $W_L W_L \to W_L W_Lh$ & 5.2 & 3.8 & 9.9 & 5.3 \\
    \hline
     $W_L W_L \to Z_L Z_Lh$ & 4.7 & 3.4 & 8.9 & 4.8 \\
    \hline
     $Z_L Z_L \to W_L W_Lh$ & 4.6 & 3.4 & 8.9 & 4.7 \\
    \hline  
    $W_L Z_L \to W_L Z_Lh$& 5.2 & 3.8 & 10  & 5.3 \\
    \hline
    \end{tabular}
    \caption{Energy scales of unitarity violation for the chosen HEFTs involving derivatives that modify the SM Lagrangian given in TeV for the $2\to 3$ processes. For the scalar singlet, $\lambda=8.3$, for the scalar doublet, $M= 45$ GeV, where $M^2= \frac{\kappa^2}{\lambda_\Sigma}$. Similarly to the $2\to 2$ case, the $hhh$ state has no unitarity violation scale from the derivative terms when the bare mass is set to zero. This corresponds to the right plot in Fig.~\ref{fig:res1}}
    \label{table:res23}
\end{table}

\bibliographystyle{JHEP} % We choose the "plain" reference style
\bibliography{main} % Entries are in the main.bib file

\end{document}